\PassOptionsToPackage{colorlinks=true,allcolors=blue}{hyperref}
\documentclass[journal=jpcafh,manuscript=article,nobib]{achemso}

\usepackage[normalem]{ulem}
\usepackage{graphicx}
\usepackage{xcolor}
\usepackage{booktabs}

\usepackage{subcaption}
\usepackage{caption}
\usepackage{xurl}
\usepackage{microtype}
\usepackage{comment}

\newcommand{\kjmol}{\mbox{kJ/mol}}

\setkeys{acs}{doi=true}

\title{Intermolecular Interactions of Large Systems: Boron Nitrides, Acenes, and Coronenes}

\author{Vladimir Fishman}
\affiliation{Department of Molecular Chemistry and Materials Science, Weizmann Institute of Science, 7610001 Re\d{h}ovot, Israel}
\author{Jan M. L. Martin}
\affiliation{Department of Molecular Chemistry and Materials Science, Weizmann Institute of Science, 7610001 Re\d{h}ovot, Israel}
\author{A. Daniel Boese}
\affiliation{Department of Chemistry, University of Graz, Heinrichstrasse 28/IV, 8010 Graz, Austria}
\email{adrian_daniel.boese@uni-graz.at}

\begin{document}
    
\begin{abstract}
In a recent contribution [Fishman et al., \textit{J. Chem. Theory Comput.} \textbf{2025}, \textit{21}, 2311–2324], we introduced another approach to benchmarking non-covalent interactions by not benchmarking interaction energies of different dimers, but by considering the evolution of interaction energies with increasing system size.

Here, we extend the benchmark set to more species, such as electrostatically bound borazine dimers as well as the minimum-energy structures of parallel displaced acene and coronene dimers.
While the minimum structures of the parallel displaced acene dimers yield similar results to previously published sandwich-structured acenes, the borazine dimers behave vastly different, yielding a yet more complete picture of non-covalent interactions and their scalability.
In contrast, the polycyclic aromatic hydrocarbon structures -- coronenes sandwich-stacked and coronenes parallel displaced -- give results consistent with those obtained for both types of the polyacene series, resulting in an updated estimate for the coronene dimer energy. By doing so, we are able to break down various approximations made by state-of-the-art Coupled-Cluster methods.

\end{abstract}

\begin{tocentry}
        \includegraphics[scale=0.31]{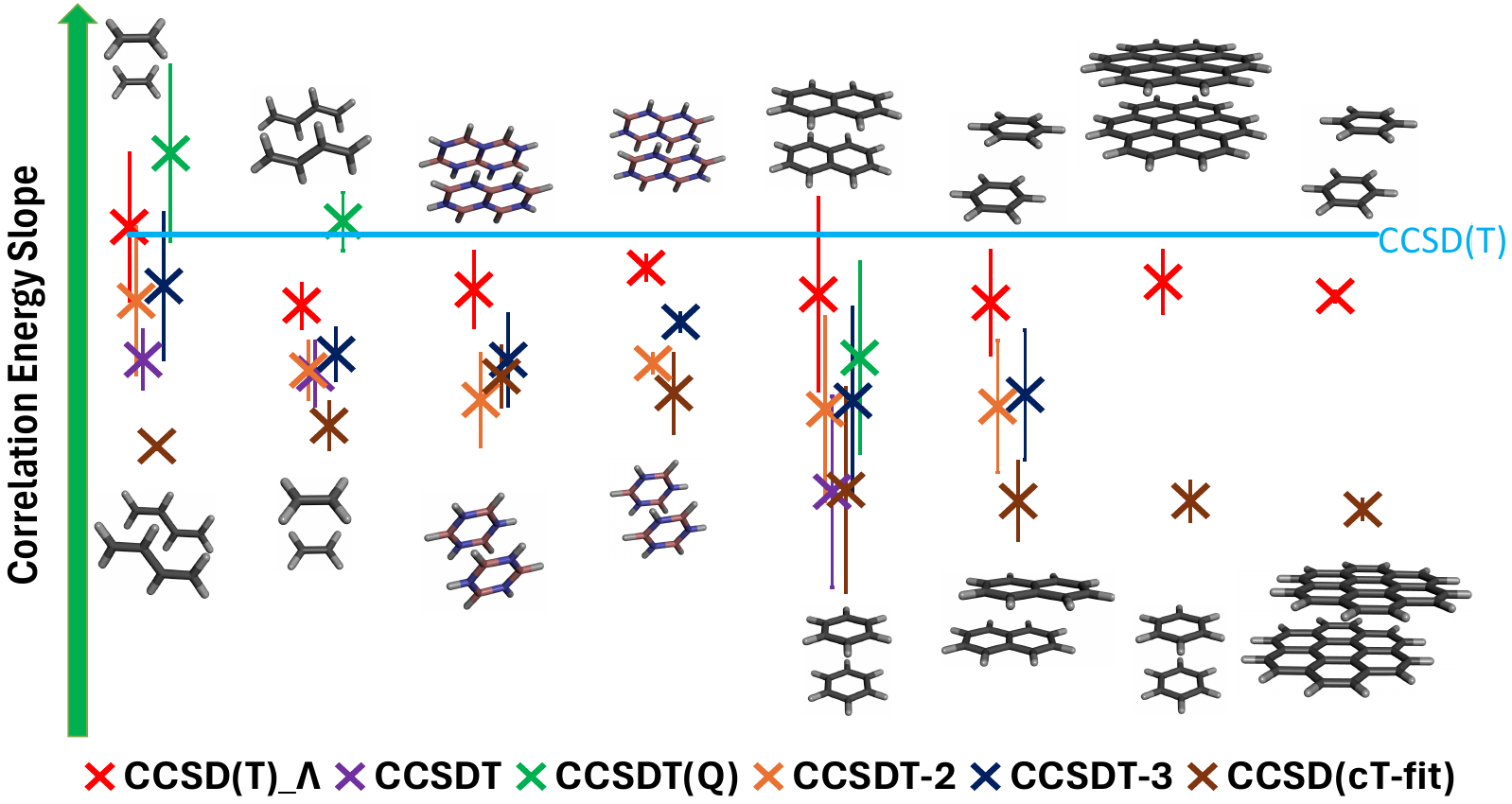}
        
        By considering the evolution of the correlation binding energy with respect to the number of subunits of dimer interactions, deviations from the reference CCSD(T) method scale linearly to ever larger systems. Overall, eight linear series are investigated.
\end{tocentry}

\flushbottom
\maketitle
\thispagestyle{empty}

\section{Introduction}
Non-covalent, also called intermolecular interactions, are at the forefront of modern science, as they determine the properties of many chemical systems, including molecular crystals \cite{saleh2012revealing, hoja2017first}, solvents \cite{storer2022surface}, polymers \cite{buaksuntear2022non, chen2022probing}, proteins and nucleotides \cite{adhav2023realm}. These interactions also control the self-assembly of nanomaterials \cite{tuncel2011non}, drug docking processes \cite{rehman2015studying}, liquids, and the binding of reaction partners in supramolecular catalysis \cite{raynal2014supramolecular}. Non-covalent interactions are weak contacts compared to covalent bonds and thus very sensitive to the environment. Therefore, their strength and nature should be predicted using robust, accurate quantum chemical methods with as few errors as possible.

In a paper from 2021,\cite{AlHamdani2021} a large discrepancy of 25\% was found between two reference methods solving the Schr\"{o}dinger equation differently,
namely fixed-node diffusion Monte-Carlo (FN-DMC) \cite{reynolds1982fixed}, which is strongly used in solid state physics (but also substantially in molecular quantum chemistry \cite{della2025fixed}), and coupled-cluster (CC) theory including singles, doubles, and perturbative triples CCSD(T) \cite{Raghavachari1989, MPICCSDpT} -- which had been termed the ``gold standard'' in gas phase quantum chemistry when utilized at its basis set limit \cite{pickard2006ccsd}. We have to note that, although CCSD(T) was employed, it was done so within its local natural orbital (LNO) approximation, investigating a series of cut-off values. The systems which showed these deviations were, however, too large for other, more elaborate methods which might yield further insights than the methods used in the Ref. \citenum{AlHamdani2021}.

We previously suggested another approach for benchmarking non-covalent interactions by considering the evolution of the correlation binding energy with respect to the number of subunits of dimer interactions.\cite{fishman2025another} We do not look at mere interaction energies, but at the deviations from the interaction energy slopes of increasing cluster sizes which scale linearly to ever larger systems. Thus, we are able to extrapolate these deviations to huge systems, specifically at the sandwich-stacked polyacene series and at two types of geometry of alkapolyene dimers, assuming the errors increase linearly as well. 

For the sandwich-stacked polyacene series, which we show below to be closely related to the coronene dimer, where some of the discrepancies between CCSD(T) and FN-DMC were found, we were able to perform extremely accurate Rank-reduced \cite{Lesiuk22} CCSDT(Q)\cite{bomble2005coupled} calculations. CCSDT(Q) is going beyond CCSD(T), taking further excitation terms in the coupled cluster approximation such as full triples and perturbative quadruples into account. CCSDT(Q) has a much steeper scaling with system size, and is also very close to the exact solution of the electronic Schr\"{o}dinger  equation for closed-shell systems without much multi-reference character.
By assuming that the CCSDT(Q) estimate is close to the near-exact solution for intermolecular interactions \cite{boese2013assesment}, we arrived in Ref. \citenum{fishman2025another} at a maximum deviation of 4.5\% from CCSD(T) and not 25\% for the buckycatcher, for which the largest discrepancy to FN-DMC and LNO-CCSD(T) had been reported in Ref.\citenum{AlHamdani2021}. Furthermore, for the coronene \emph{parallel displaced} dimer, we arrived at a maximum deviation of 3.5\% between CCSDT(Q) and CCSD(T) and not the 12\% difference between LNO-CCSD(T) and FN-DMC.
Summarizing, whereas CCSD(T) indeed tends to overestimate the dissociation energies of pi-stacks, it does not so with the large magnitude suggested by FN-DMC.
Similar results have been also obtained for several polyaromatic hydrocarbon stacks including benzene using a simple underlying PPP Hamiltonian;\cite{Lambie2025CCSDTDispersion} here, however, the CCSDT(Q) interaction energies exhibit larger dissociation energies than CCSD(T), which is in contrast to the results when a HF wavefunction is utilized for CCSDT(Q).

Despite these results, all polyacenes and polyaromatic hydrocarbon stacks were investigated in our last contribution\cite{fishman2025another} at their sandwich, \textit{i.e.} transition state structures because we were able to utilize their high symmetry.\cite{briccolani2026weak}
 One may argue that especially higher-order dispersion terms might be canceled at high symmetries \cite{Semidalas2025}, whereas at lower symmetries, they might not, which is also discussed in Ref. \citenum{herbert2021neat}. For this purpose, we investigate \texttt{parallel-displaced} polyacene dimers which are global minimum structures for the polyacene series for naphthalene dimer and larger, and nearly isoenergetic\cite{Hobza1996BenzeneDimerIsoenergetic} with the slanted T-shaped local minimum structure \cite{law1984dimers} for the benzene dimer. Other types of geometries of polyacenes, specifically glike, cross and X, have been described in the most recent paper of Sharapa and co-workers.\cite{matsokin2026revisiting} Additionally, we explore sandwich-structured and parallel displaced coronene dimers. Furthermore, we consider two sandwich-like borazine dimer structures: One where the BN moiety is stacked towards NB, which exhibits a large electrostatic attraction, and the other one with the BN moiety stacked towards BN, yielding electrostatic repulsion.

This contribution is hence a direct continuation of our previous paper\cite{fishman2025another}, introducing new species with (essentially perfect) lines to interpolate. By using DFT-based symmetry adapted perturbation theory (DFT-SAPT), we distinguish these species and groups them into different categories. 

\section{Computational Details}

Calculations of the electronic structure have been performed using various software packages, depending on the method employed.
Geometries of the investigated systems in the paper have been optimized by density functional theory (DFT)\cite{Hohenberg1964,Kohn1965} utilizing the $\omega$-B97M-V \cite{Mardirossian2015} functional with the def2-QZVPP\cite{Weigend2005} basis set (here, we omitted diffuse functions because of linear dependencies) using ORCA  5.0.3.\cite{neese2020orca} Full molecular symmetry has imposed.

To achieve highly accurate single-point energy calculations for the optimized structures, various forms of Coupled-Cluster (CC) theory have been employed:
\begin{enumerate}

   \item The ``gold-standard'' canonical CCSD(T) (Coupled-Cluster with Single, Double, and Perturbative Triple Excitations)\cite{Raghavachari1989,MPICCSDpT}, including Resolution of Identity (RI), also called Density Fitting (DF), \cite{epifanovsky2013general,shen2019massive} for large systems, performed by the MRCC\cite{MRCC} program package (August 2023 and later versions).
   
   \item LNO (Local Natural Orbitals)-CCSD(T)\cite{Nagy2019,local_corr_review2024} using the Normal, Tight, VeryTight (denoted vTight from hereon) and VeryVeryTight (vvTight) threshold settings with localcc=2021.
   
   \item DLPNO(Domain-Based Local Pair Natural Orbital)-CCSD(T$_1$)\cite{riplinger2013efficient,riplinger2013natural,riplinger2016sparse,guo2018communication} using the TightPNO threshold setting with two different values for the Pair Natural Orbital (PNO) truncation (TCutPNO), in which orbitals with an occupation number less than TCutPNO are neglected. The TCutPNO parameters used are $10^{-6}$ and $10^{-7}$, respectively, and DLPNO-CCSD(T$_1$) calculations are done with ORCA 5.0.4.\cite{neese2020orca}
  
   \item PNO(Pair Natural Orbital Local)-CCSD(T)\cite{ma2018explicitly} with Domopt Tight, Domopt vTight, and Pairopt Tight truncation settings, performed by means of MOLPRO 2024.3.\cite{Werner2020}
  
   \item The CCSD(T)$_\lambda$ method\cite{stanton1996simple, crawford1998investigation, kucharski1998noniterative, kucharski1998sixth} --- which treats the triples as a perturbation to CCSD --- performed by the CFOUR program software.\cite{CFOUR} (On a side note, the CCSD(T) equations can be derived \cite{stanton1997ccsd} by approximating the $\Lambda$ vector as the transpose of the doubles amplitudes vector $T_2$.)

   \item Post-CCSD(T) methods, specifically CCSDT-2 and CCSDT-3 \cite{noga1987towards}, for the species with maximum two aromatic rings, were performed using CFOUR\cite{CFOUR}. Both methods go beyond CCSD(T) by introducing iterative $T_3$, and the first formal increment beyond CCSD(T) is the fifth-order TQ term. Then, the difference between CCSDT-2 and CCSDT-3 appears only at the sixth order of perturbation theory. It arises from the contribution of the effect on the $T_3$ amplitudes equation of single excitations, which is neglected in CCSDT-2 but included in CCSDT-3 (See Ref. \citenum{He2001comparisonCCSDTs} for more details). 

\end{enumerate}

For more approximate single-point energy calculations, we have employed WFT (wavefunction theory)-based methods, including MP2 (second-order Møller–Plesset perturbation theory)\cite{moller1934note} using the MRCC\cite{MRCC} software and MP3 (third-order Møller–Plesset perturbation theory) with its RI implementation\cite{Loipersberger2021} in the Q-Chem 6.1.0 program package.\cite{Shao2015} Additionally, DFT-based methods are utilized, specifically the dRPA (direct random-phase approximation) as implemented in MRCC\cite{MRCC} (August 2023) and DFT-SAPT(symmetry-adapted perturbation theory) as available in MOLPRO\cite{Werner2020} 2022.2. These methods provide a lower-cost alternative for estimating correlation effects while maintaining a reasonable level of accuracy.

All single-point calculations have been performed using the correlation-consistent\cite{dunning1989gaussian} and augmented correlation-consistent\cite{woon1993gaussian} polarized valence X-zeta (X = D, T, Q, 5) (aug-)cc-pVXZ basis sets. These will be abbreviated henceforth with DZ, TZ, QZ, 5Z, as well as aDZ, aTZ, aQZ, and a5Z, respectively.

\section*{Results and Discussion}

\subsection{Strong Correlation Diagnostics}
To determine if our systems are indeed well-described by single determinant methods, we evaluate a strong correlation diagnostic\cite{martin2022exchange} for all series based on total atomization energy (TAE) calculations obtained with the TZ basis set using the TPSS\cite{TPSS-functional} functional:
\begin{equation}
\label{eq:TPSS-CCSDt}
    \% TAE_X[\mathrm{KS-HF(TPSS)}]=100 \% \times \frac{(\mathrm{TAE_X[TPSS@TPSS]-TAE_X[TPSS@HF]}}{TAE[\mathrm{CCSD(T)}]}
\end{equation}
Thus, the correlation diagnostic is calculated by the difference in the total atomization energies between the TPSS exchange using a TPSS exchange wavefunction ($TAE_X[TPSS@TPSS]$) and TPSS exchange using a Hartree-Fock exchange wavefunction ($TAE_X[TPSS@HF]$) divided by the total atomization energy of CCSD(T).
The results obtained by eq. \ref{eq:TPSS-CCSDt} show small static correlation around 5\% $\% TAE_X$ for polyacene PD series and negligible static correlation, less than 2\%, for borazine series, which, most importantly and interestingly, do not become much larger when the number of rings increases. Another ($TAE_X[(T)]$) diagnostic from Ref. \citenum{martin2022exchange} (Table 1, Cluster 3) confirms these observations.
 
Moreover, commonly used rule of thumb for the default $T_1$ and $D_1$ diagnostics, $\leq 0.02$ \cite{lee1989diagnostic, lee1989theoretical} and $\leq 0.05$ \cite{langhoff1988ab} respectively, reaffirm that single-determinant reference should provide adequate results for a single-reference coupled-cluster treatment for all investigated in our study systems.

Meanwhile, the HOMO-LUMO gap decreases progressively along the number of subunit.

For more details read SI on pages 81-87.

\subsection{Different species investigated}
In order to distinguish the different species, we explore the type of their intermolecular interactions by the use of symmetry-adapted perturbation theory, in which the total energy is expanded as
\begin{equation}
E_{\rm total}=E_{\rm elec}^{(1)}+E_{\rm ex}^{(1)}+E_{\rm ind}^{(2)}+E_{\rm ind-ex}^{(2)}+E_{\rm disp}^{(2)}+E_{\rm disp-ex}^{(2)}+E_{\rm \Delta HF}
\end{equation}
The induction $E_{\rm ind}=E_{\rm ind}^{(2)}+E_{\rm ind-ex}^{(2)}+E_{\rm \Delta HF}$ and  electrostatic $E_{\rm elec}^{(1)}$ energies have been obtained from DFT-SAPT and their ratios to dispersion $E_{\rm disp}=E_{\rm disp}^{(2)}+E_{\rm disp-ex}^{(2)}$ \cite{rezac2011s66}, displayed in Table \ref{tab:DFTSAPT}, have been found as most instructive when trying to separate dispersion-dominated from induction-dominated systems.


A more detailed inspection would analyze the percentage of the mean contribution of a fitted leading-order dispersion $C_6$ coefficient in the SAPT dispersion contribution, which is very small for induction-dominated systems.\cite{Masumian2023}

\begin{table}[!htbp]
\caption{Different DFT-SAPT components (\kjmol), induction to dispersion, and electrostatic to dispersion ratios using the cc-pV5Z basis set.}
\label{tab:DFTSAPT} 
\centering
{\footnotesize
\begin{tabular}{lccccc|cc}
\hline\hline
        Dimer & $E_{\rm total}$ & $E_{\rm elec}^{(1)}$ & $E_{\rm ex}^{(1)}$ & $E_{\rm ind}$ & $E_{\rm disp}$ & $\frac{E_{\rm ind}}{E_{\rm disp}}$ &  $\frac{E_{\rm elec}^{(1)}}{E_{\rm disp}}$ \\ \hline
Benzene (sandwich) & -7.4 & 0.3 & 12.6 & -0.8 & -19.6 & 0.04 & -0.02 \\ 
Benzene (parallel displaced) & -11.8 & -5.8 & 23.2 & -2.6 & -26.8 & 0.10 & 0.22 \\ 
Ethylene & -0.4 & 1.1 & 1.0 & -0.1 & -2.4 & 0.04 & -0.46 \\
Borazine (syn) & -7.8 & -2.0 & 15.2 & -0.5 & -20.5 & 0.02 & 0.10 \\
Borazine (anti) & -13.0 & -16.3 & 38.7 & -2.2 & -33.3 & 0.06 & 0.49 \\ \hline
\\ \hline
Naphthalene (sandwich) & -17.1 & -3.0 & 28.1 & -1.3 & -40.8 & 0.03 & 0.07 \\ 
Naphthalene (parallel displaced) & -26.0 & -15.5 & 50.6 & -5.2 & -56.0 & 0.09 & 0.28 \\ 
Coronene (sandwich)           & -57.8 & -20.1 & 87.9  & -1.9 & -123.2 & 0.02 & 0.16 \\
Coronene (parallel displaced) & -78.1 & -36.6 & 114.1 & -3.0 & -145.3 & 0.02 & 0.25 \\ 
Trans-Butadiene (relaxed) & -3.4 & 1.3 & 6.3 & -0.5 & -10.5 & 0.05 & -0.12 \\
Trans-Butadiene (fixed) & -2.5 & 1.8 & 1.7 & -0.2 & -5.7 & 0.04 & -0.31 \\
BN Naphthalene (syn) & -14.4 & -4.3 & 29.5 & -0.8 & -38.9 & 0.02 & 0.11 \\
BN Naphthalene (anti) & -25.3 & -36.4 & 80.7 & -4.8 & -65.0 & 0.07 & 0.56 \\\hline
\end{tabular}
}
\end{table}

Based on the ratios reported in Table \ref{tab:DFTSAPT}, four distinct types of system can be distinguished.
As noted in Ref. \citenum{fishman2025another}, the $\pi$-stacks and alkapolyene stacks behave quite differently when compared to each other. This can be seen especially in the electrostatic to dispersion ratio, which is negative for the ethylene and trans-butadiene dimers, while it is positive for benzene and naphthalene (with the exception of sandwiched benzene, which is very slightly negative). Since benzene and naphthalene have sizable quadrupole moments, this is visible in the electrostatic interaction between these species. The borazine and BN-naphthalene dimers are distinctly different again, as a partial charge can be found on the nitrogen (negative) and boron (positive) atoms. This is especially visible in the anti configuration, where the largest electrostatic contribution of all dimers has been found. The syn configurations exhibit a more repulsive interaction which is, however, in its electrostatic component countered by the quadrupole and higher moments, making this configuration different again. We can already hypothesize that the alkapolyene stacks, $\pi$-stacks, borazine stacks (syn), and borazine stacks (anti) all behave quite differently, giving some insight about the nature of their intermolecular interactions. 

\subsection{Basis Set Effects}
\begin{figure}[htbp]
\centering

\begin{minipage}[t]{0.49\textwidth}
\centering
\begin{subfigure}[t]{\linewidth}
  \centering\includegraphics[width=\linewidth]{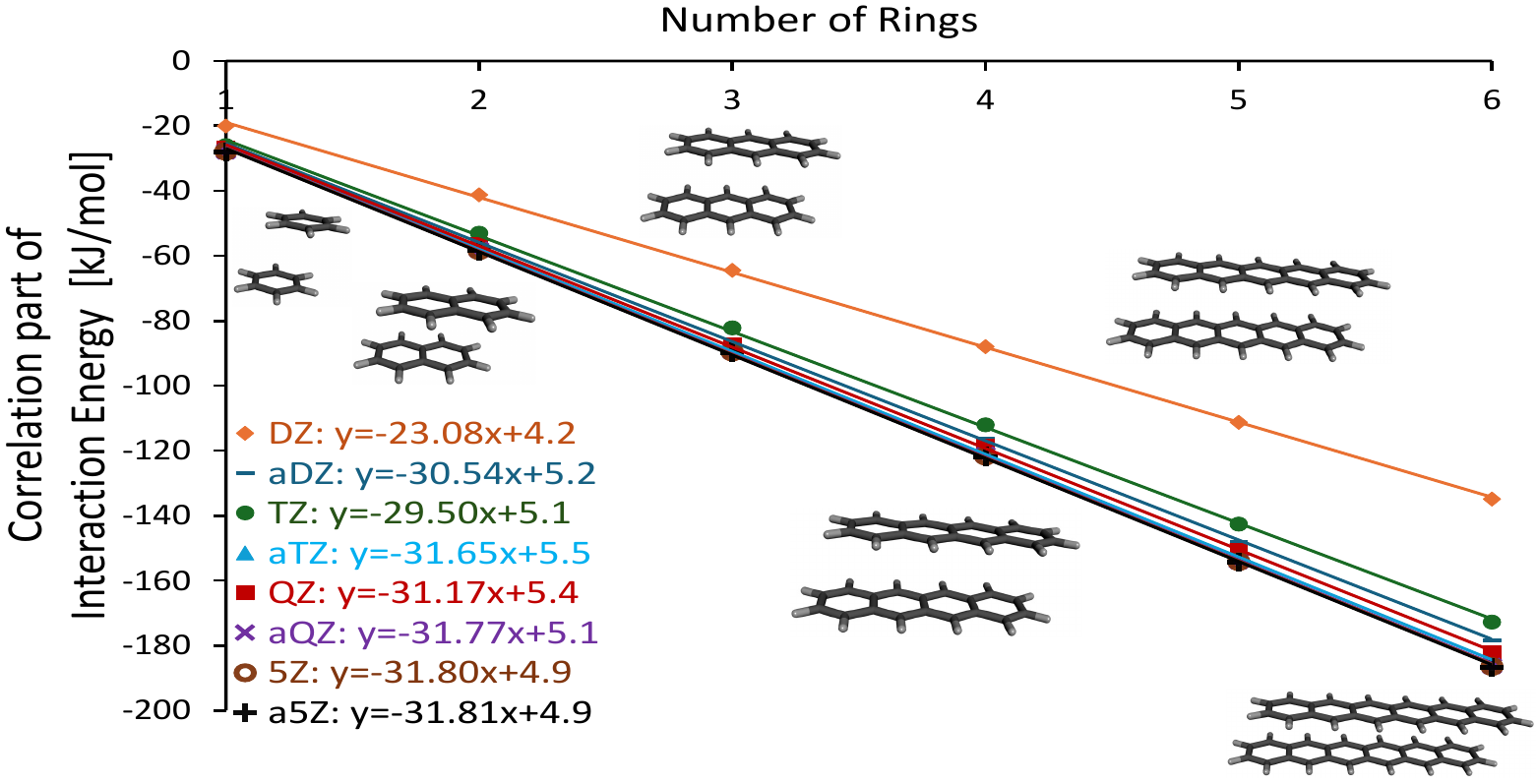}
  \caption{cp-corr energies for acenes PD}
\end{subfigure}

\medskip

\begin{subfigure}[t]{\linewidth}
  \centering\includegraphics[width=\linewidth]{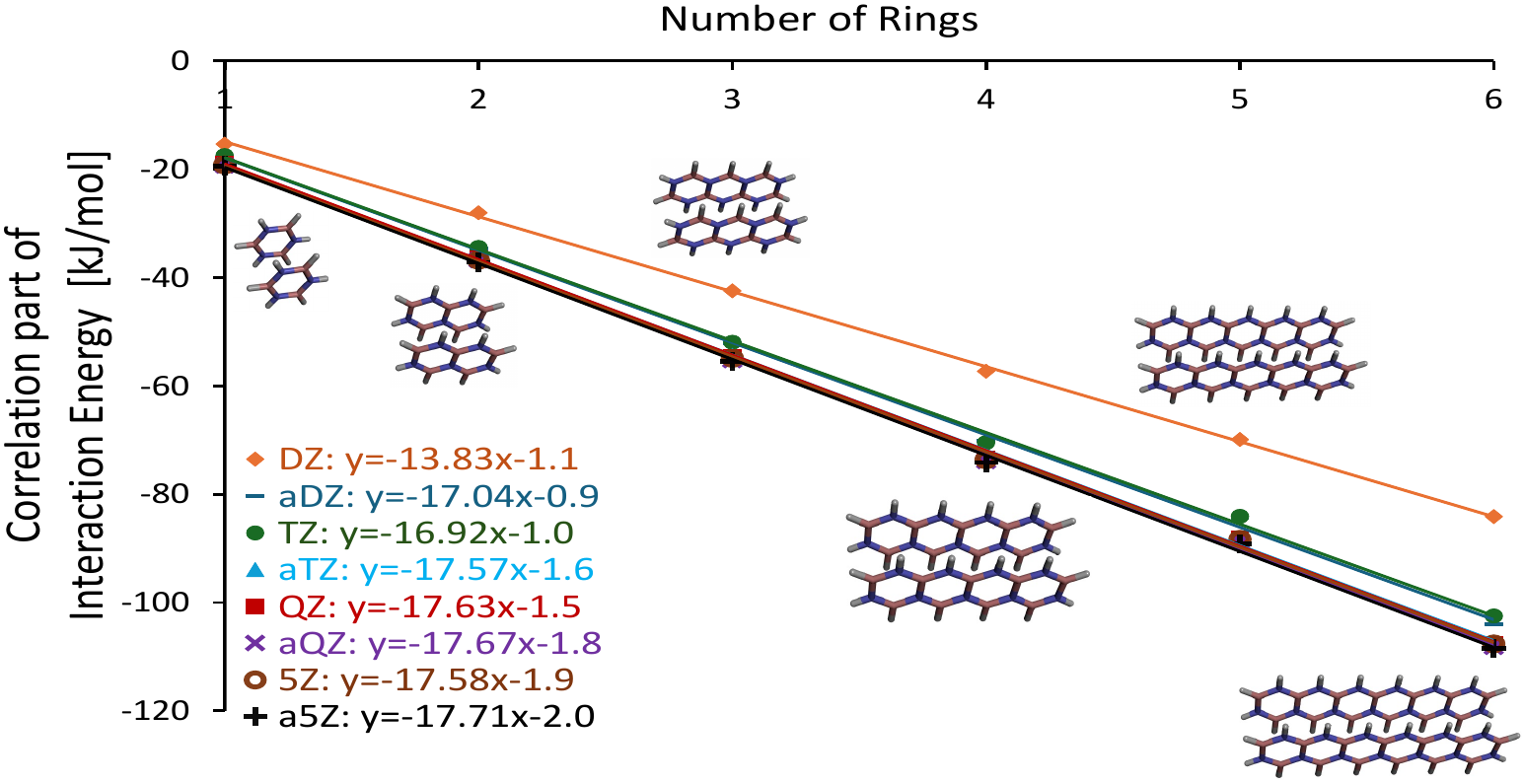}
  \caption{cp-corr energies for BNx $\cdots$ NBx}
\end{subfigure}

\medskip

\begin{subfigure}[t]{\linewidth}
  \centering\includegraphics[width=\linewidth]{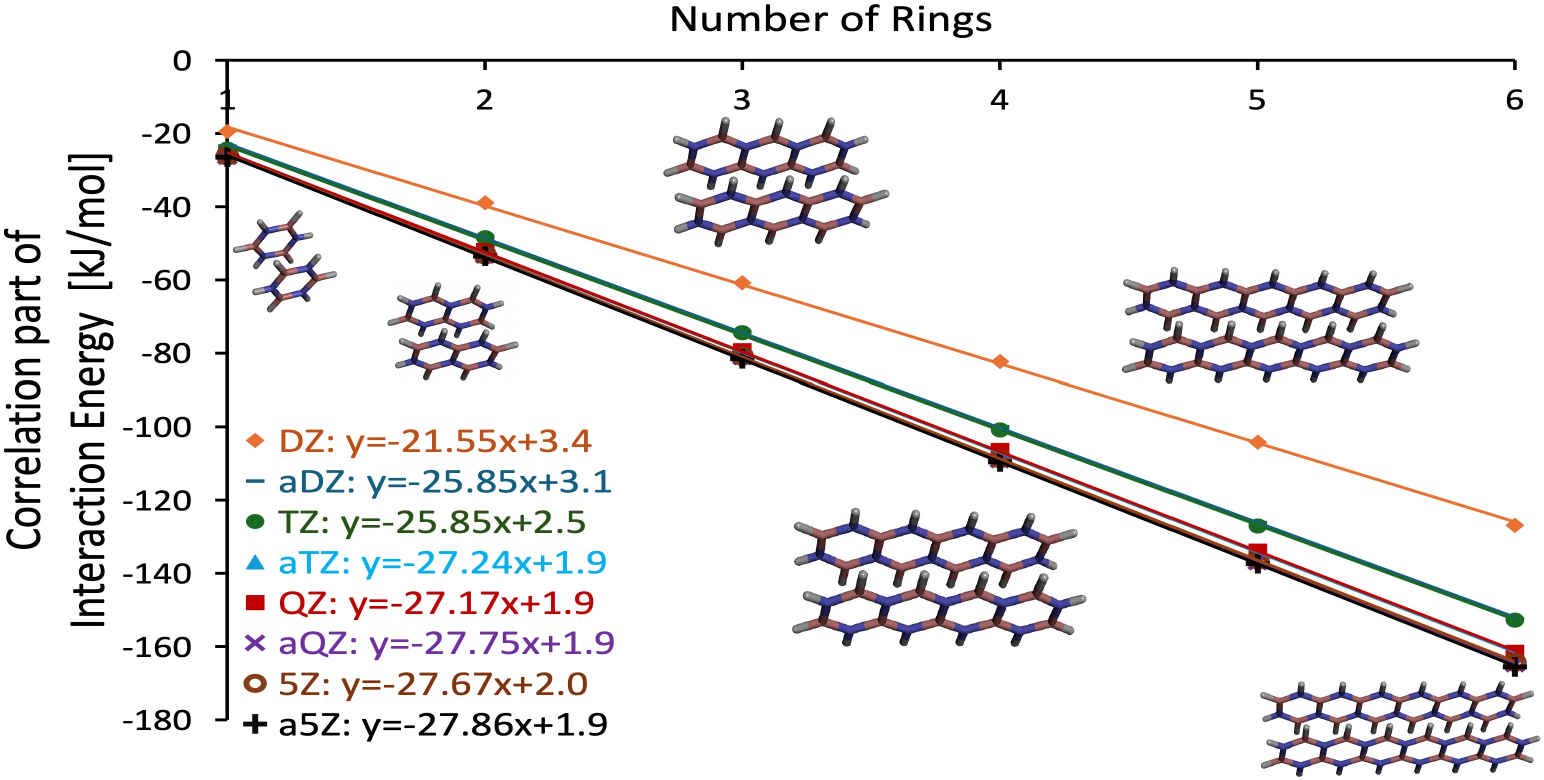}
  \caption{cp-corr energies for BNx $\cdots$ BNx}
\end{subfigure}
\end{minipage}
\hfill
\begin{minipage}[t]{0.49\textwidth}
\centering
\begin{subfigure}[t]{\linewidth}
  \centering\includegraphics[width=\linewidth]{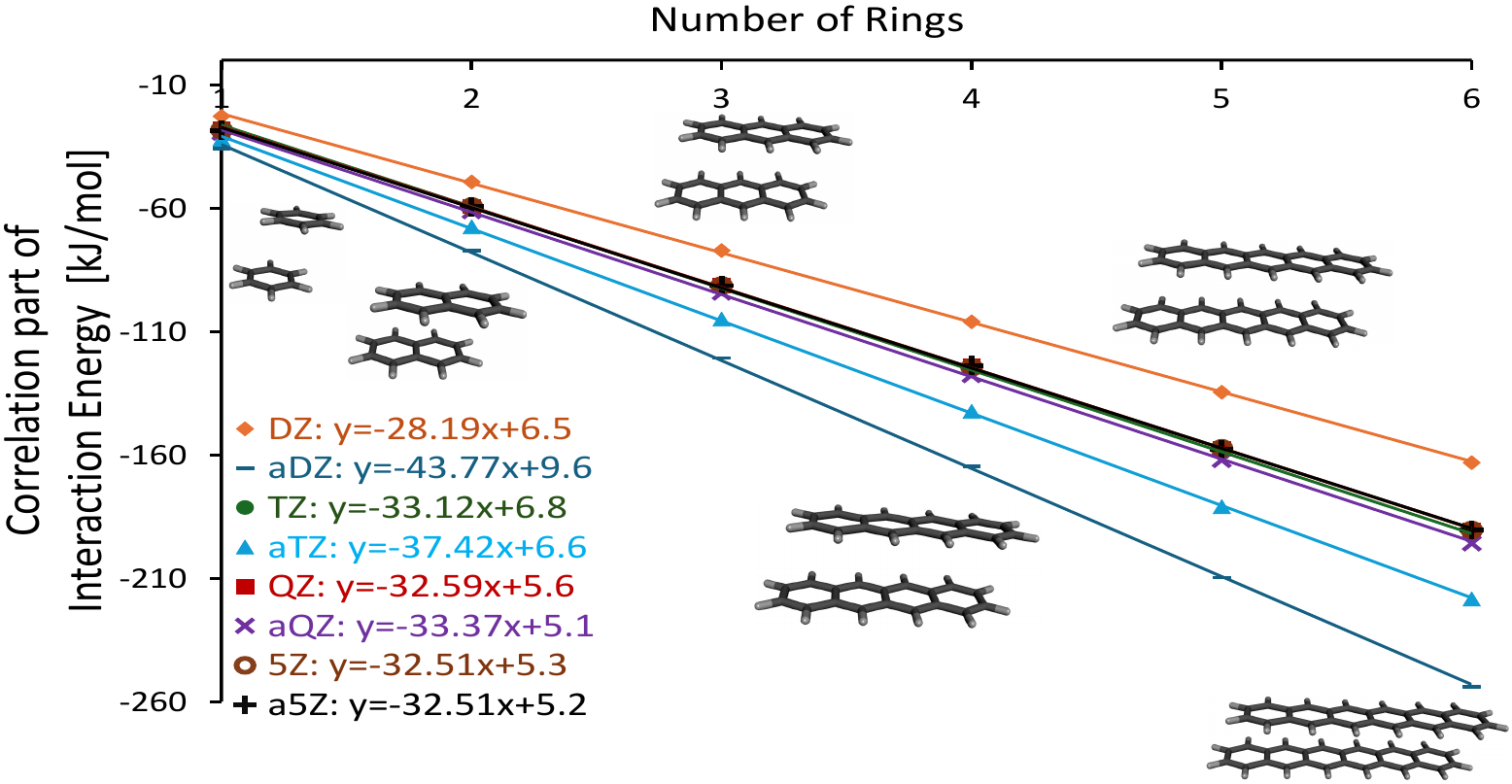}
  \caption{cp-uncorr energies for acenes PD}
\end{subfigure}

\medskip

\begin{subfigure}[t]{\linewidth}
  \centering\includegraphics[width=\linewidth]{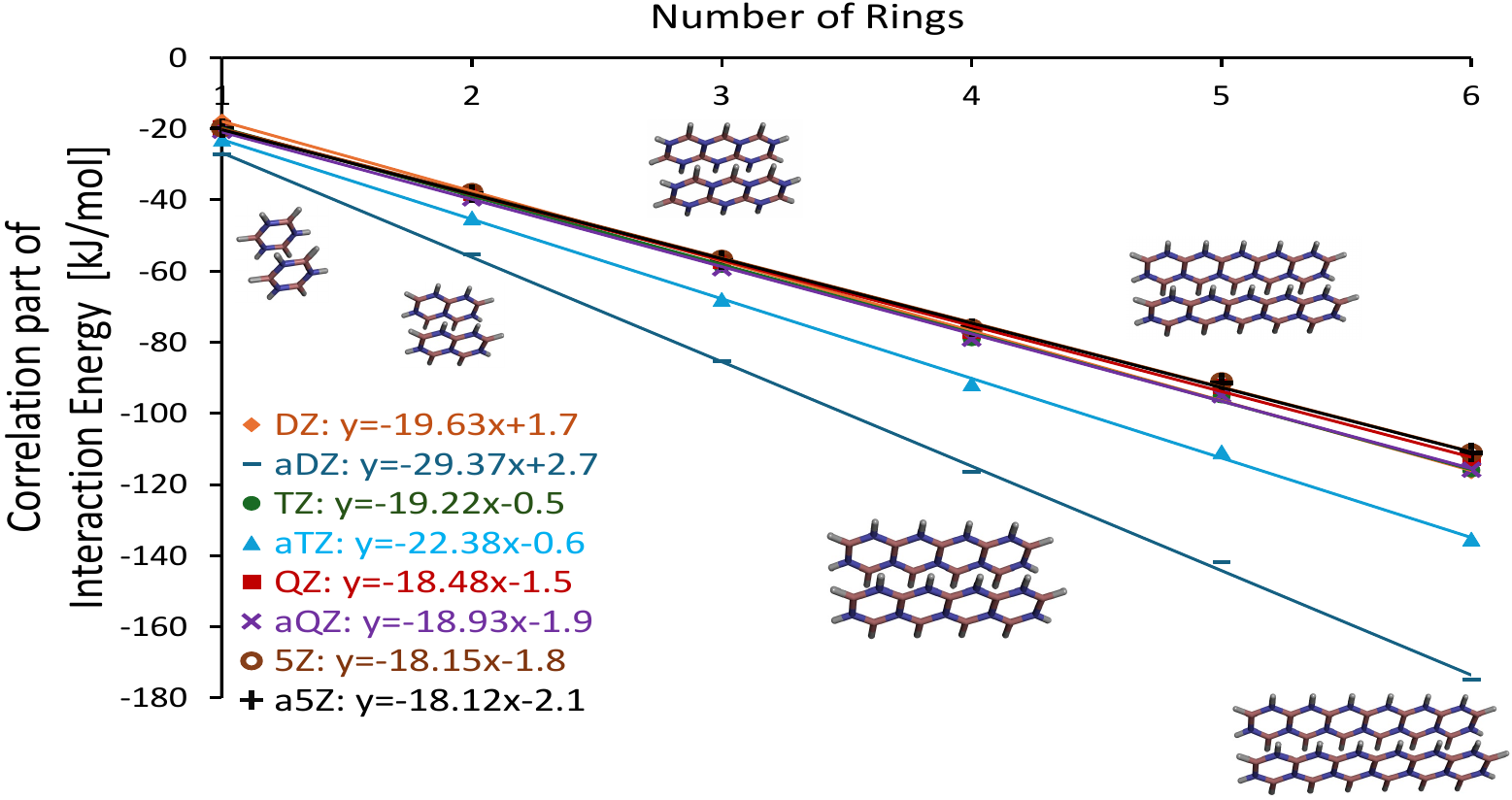}
  \caption{cp-uncorr energies for BNx $\cdots$ NBx}
\end{subfigure}

\medskip

\begin{subfigure}[t]{\linewidth}
  \centering\includegraphics[width=\linewidth]{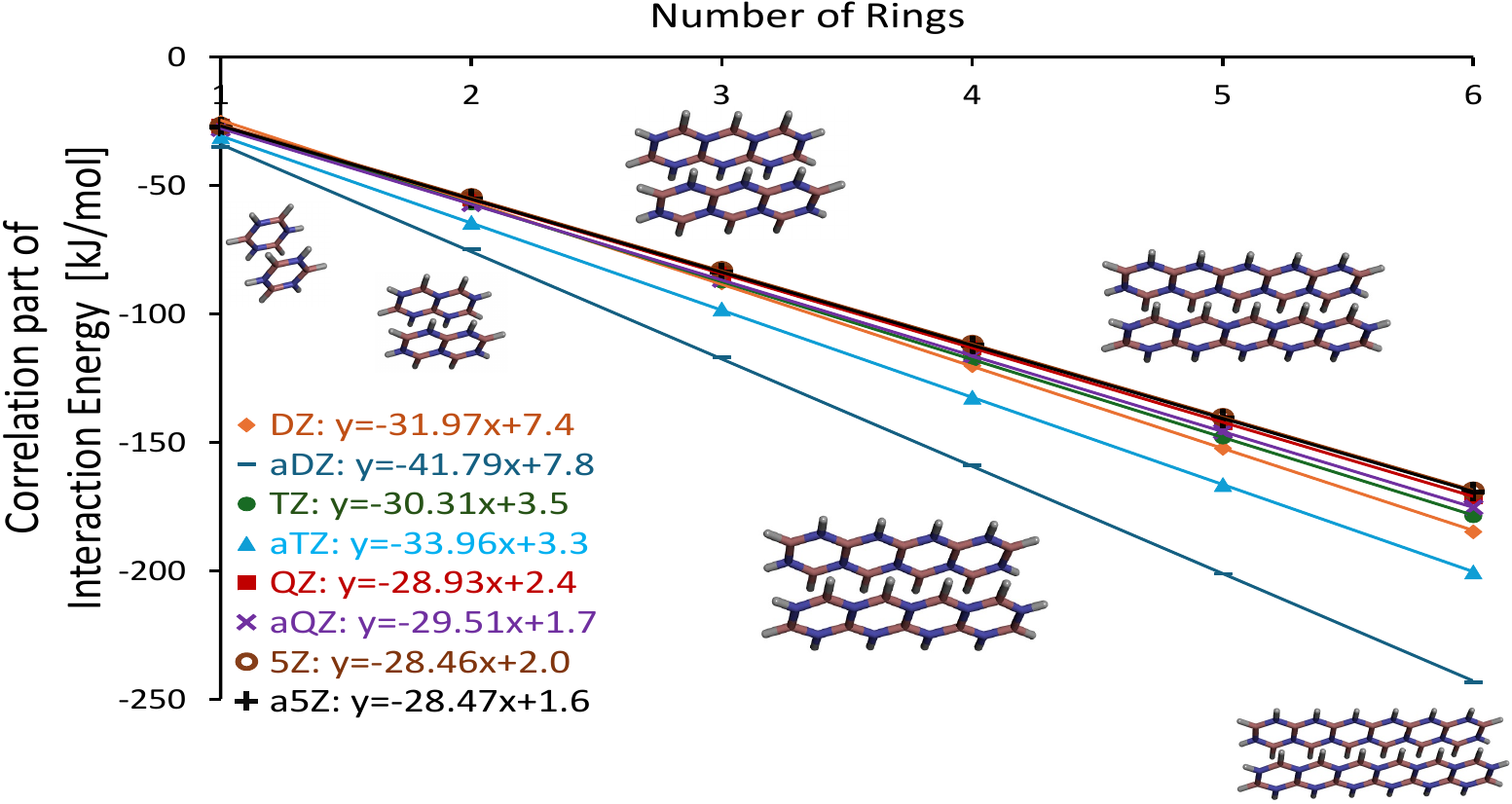}
  \caption{cp-uncorr energies for BNx $\cdots$ BNx}
\end{subfigure}
\end{minipage}

\caption{Correlation energies in \kjmol\  vs. number of rings using the cc-pV$n$Z and aug-cc-pV$n$Z ($n$=D-5) basis sets at the LNO-CCSD(T) level with Tight threshold including MP2 corrections.}

\label{fgr:BSE}

\end{figure}

In order to evaluate different basis sets, we compare results obtained at the LNO-CCSD(T) level of theory with a Tight threshold. These are probably best suited, as calculations for 6-ring systems are feasible for this method even in the a5Z basis set.
The correlation part of $\pi$-stacked molecules scales linearly with system size, yielding straight lines with $R^2>0.99$ (Figure \ref{fgr:BSE}). 

Counter-poise (cp) corrected and counter-poise uncorrected slopes of each series have different behavior of convergence to the slope that obtained from complete basis set limit noncovalent interaction energies. For instance, despite that both cp-corr and cp-uncorr \{Q,5\}Z slopes of parallel displaced polyacene series give the same result, cp-corr $n$Z slopes do not converge smoothly to this result, whereas  cp-corr $n$Z slopes do. In contrast, in BNx $\cdots$ NBx series, we can observe an opposite situation: cp-corr slopes have convergence to \{Q,5\}Z while cp-uncorr slope do not have. For BNx $\cdots$ BNx we see a highly unusual observation: cp-corr $n$Z and a$n$Z slopes converge towards to cp-uncorr \{Q,5\}Z and vice versa, cp-uncorr $n$Z and a$n$Z slopes converge towards to cp-corr \{Q,5\}Z.

Intercepts are much more sensitive to the basis set in all types of investigated systems, especially cp-uncorrected intercepts. Interestingly, the aDZ cp-uncorrected intercepts are even worse than the cp-uncorrected DZ ones (see subfigures d,e,f of Figure \ref{fgr:BSE}). Thus, for the aDZ basis set, it seems to be some imbalance introduced by the diffuse functions. 

Based on Figure \ref{fgr:BSE} a-c and corresponding cp-corrected slopes, we can summarize that the QZ basis set appears as the best compromise to get accurate slope values for a larger series of parallel displaced polyacenes and BNx $\cdots$ BNx. Practically, from a computational cost point of view, even the TZ basis set slopes are satisfactory enough, because the deviation cp-corrected TZ slope from the cp-corrected \{Q,5\}Z slope is only 5-8\% (namely: 5\% for polyacene PD dimers and for BNx $\cdots$ NBx, and 8\% for BNx $\cdots$ BNx), but to reach the sub-kJ/mol accuracy, further corrections are required.  
Optimally, if the computing resource allows, it is better to use QZ because its slope deviation to CBS is less than 4\% for all systems.


\subsection{Local Correlation}
\begin{figure}[htbp]
\centering

\begin{minipage}[t]{0.49\textwidth}
\centering
\begin{subfigure}[t]{\linewidth}
  \centering\includegraphics[width=\linewidth]{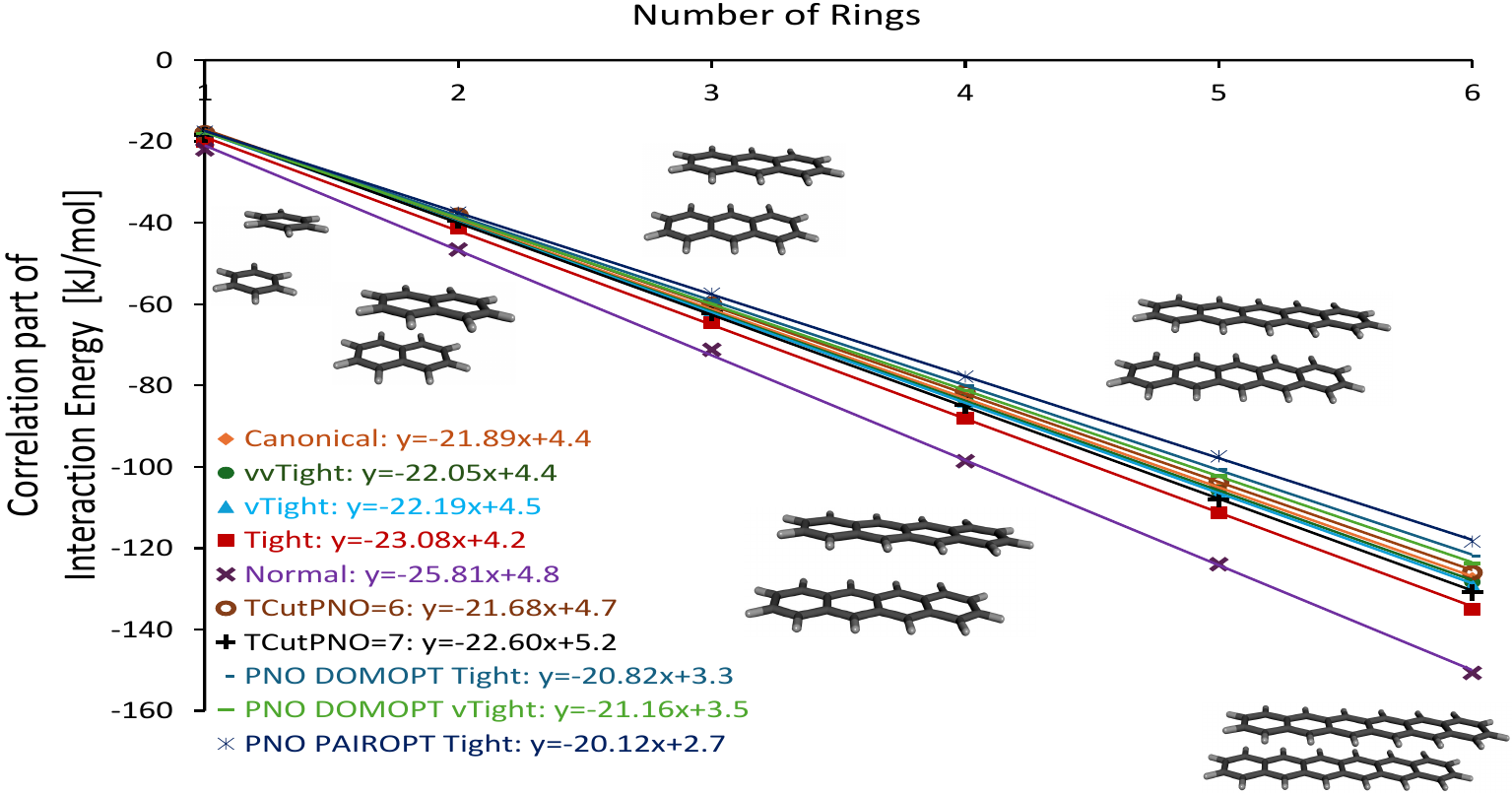}
  \caption{cp-corr DZ energies for acenes PD}
\end{subfigure}

\medskip

\begin{subfigure}[t]{\linewidth}
  \centering\includegraphics[width=\linewidth]{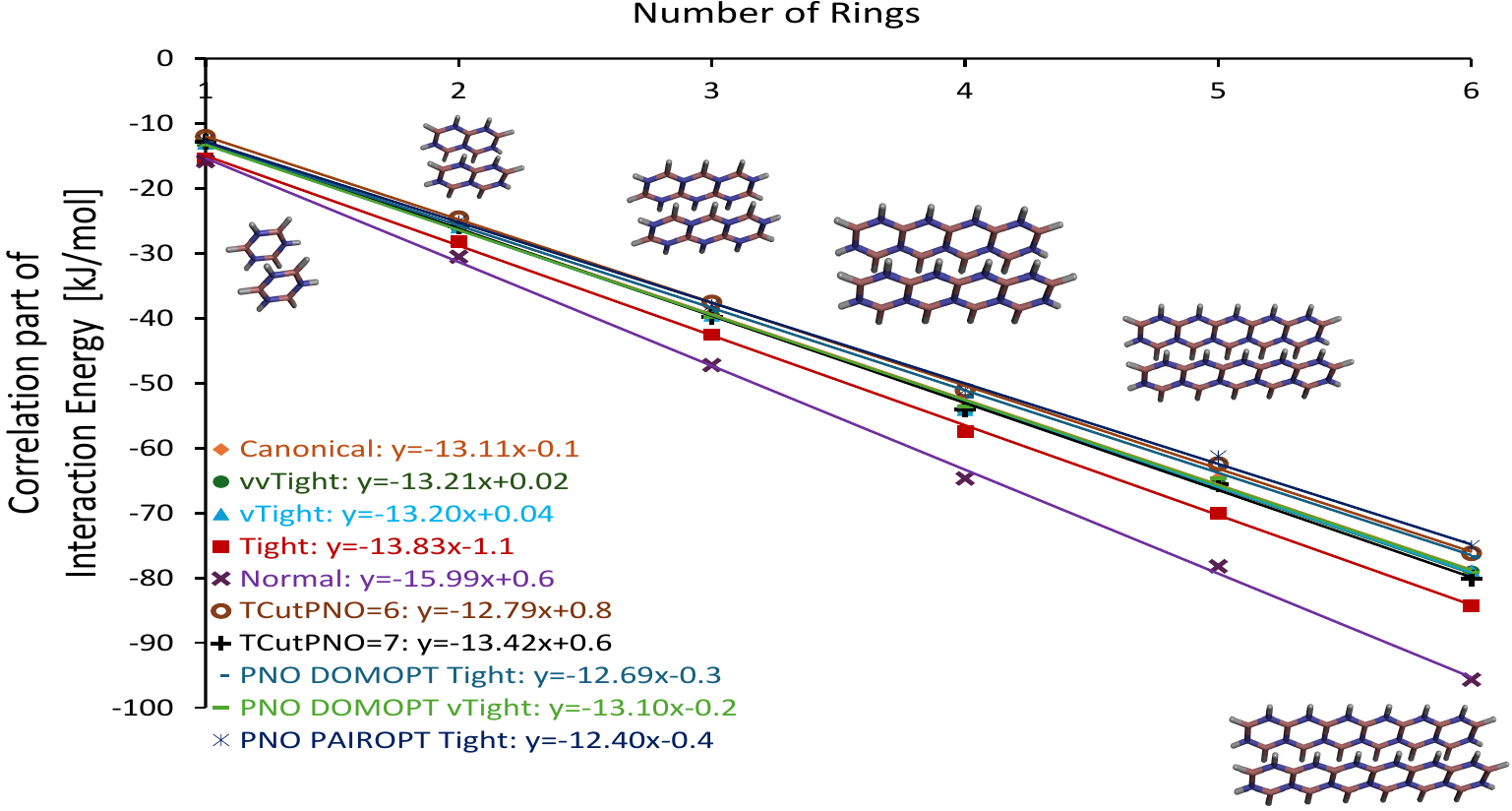}
  \caption{cp-corr DZ energies for BNx $\cdots$ NBx}
\end{subfigure}

\medskip

\begin{subfigure}[t]{\linewidth}
  \centering\includegraphics[width=\linewidth]{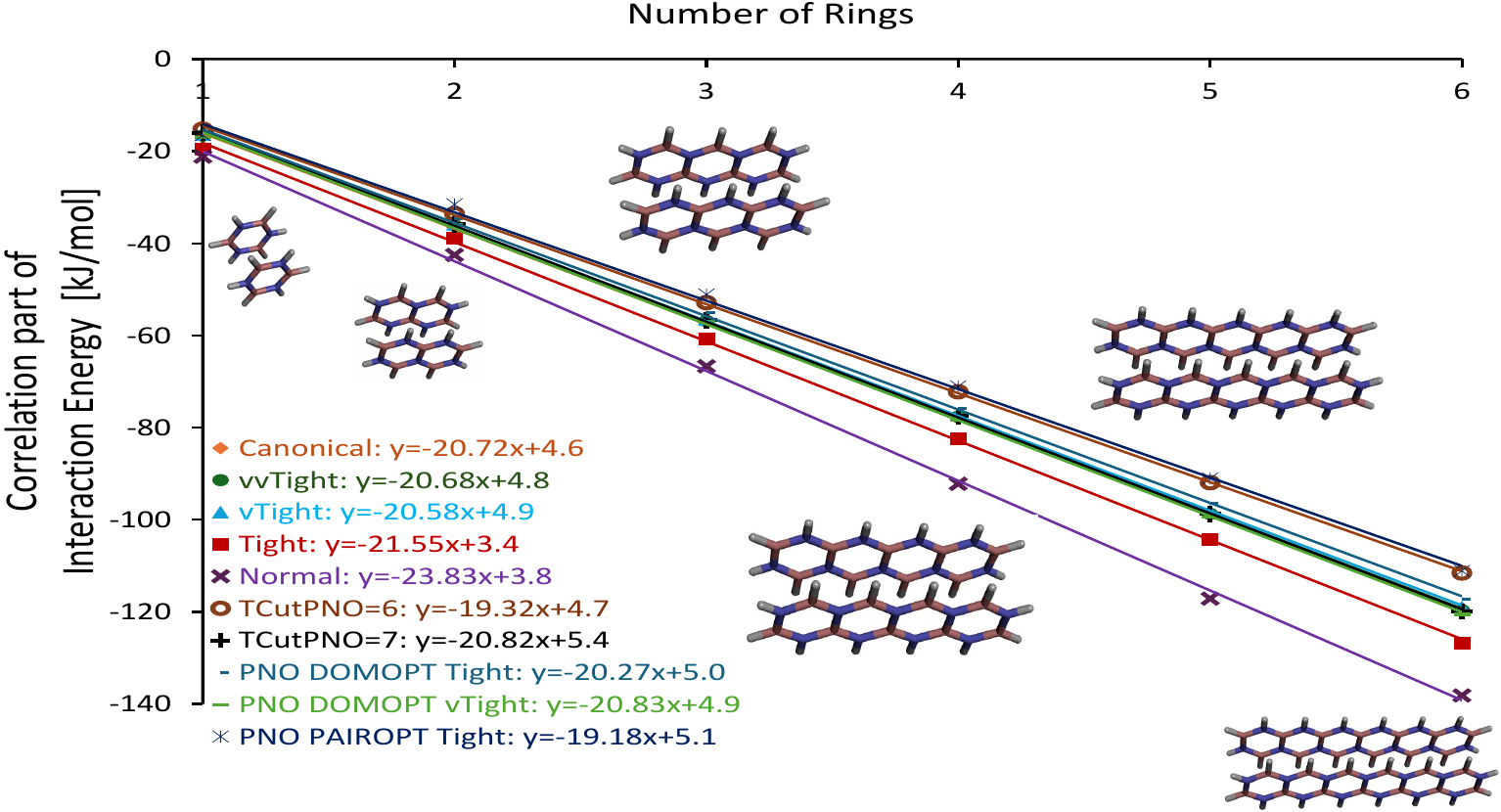}
  \caption{cp-corr DZ energies for BNx $\cdots$ BNx}
\end{subfigure}
\end{minipage}
\hfill
\begin{minipage}[t]{0.49\textwidth}
\centering
\begin{subfigure}[t]{\linewidth}
  \centering\includegraphics[width=\linewidth]{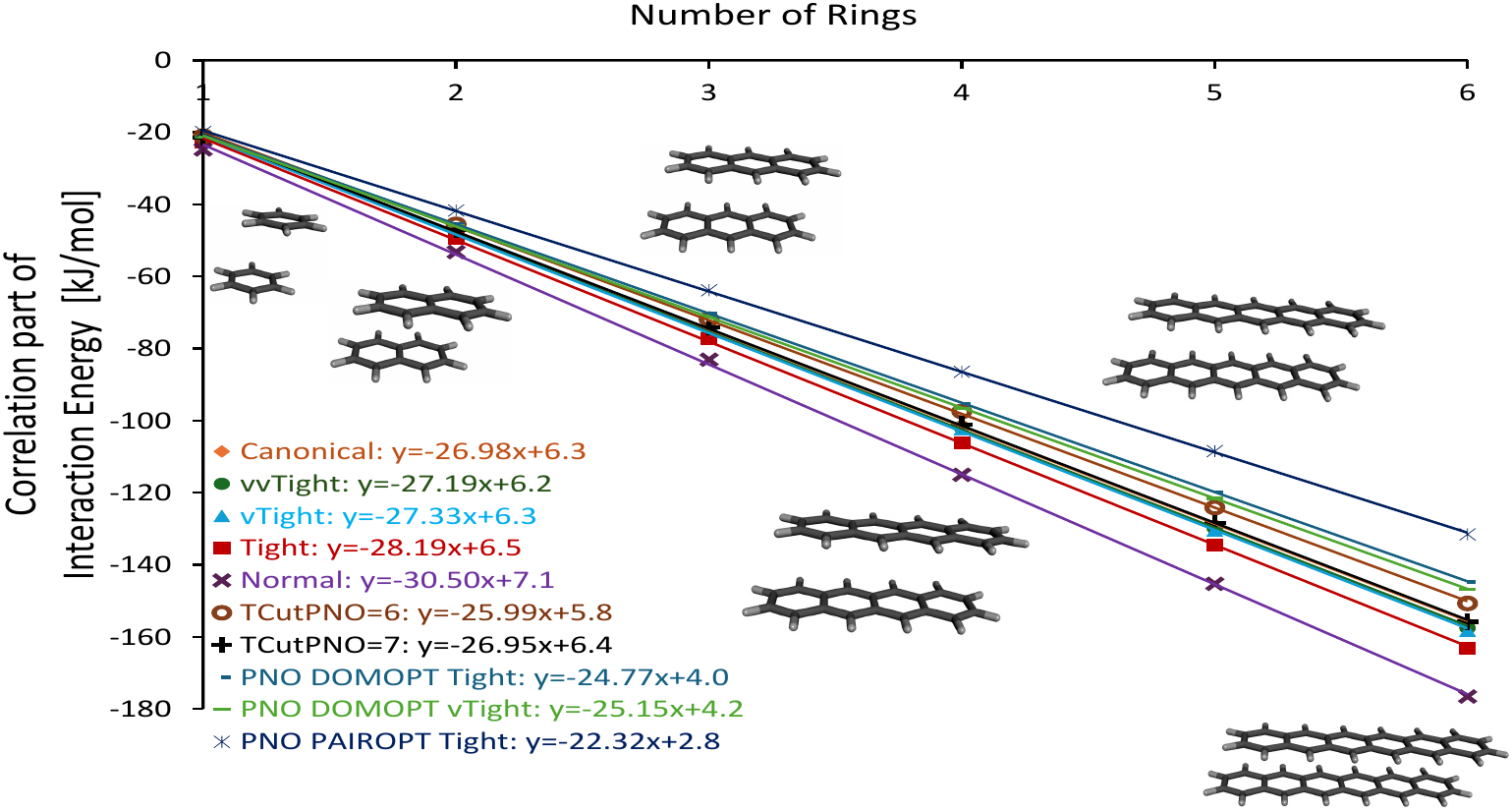}
  \caption{cp-uncorr DZ energies for acenes PD}
\end{subfigure}

\medskip

\begin{subfigure}[t]{\linewidth}
  \centering\includegraphics[width=\linewidth]{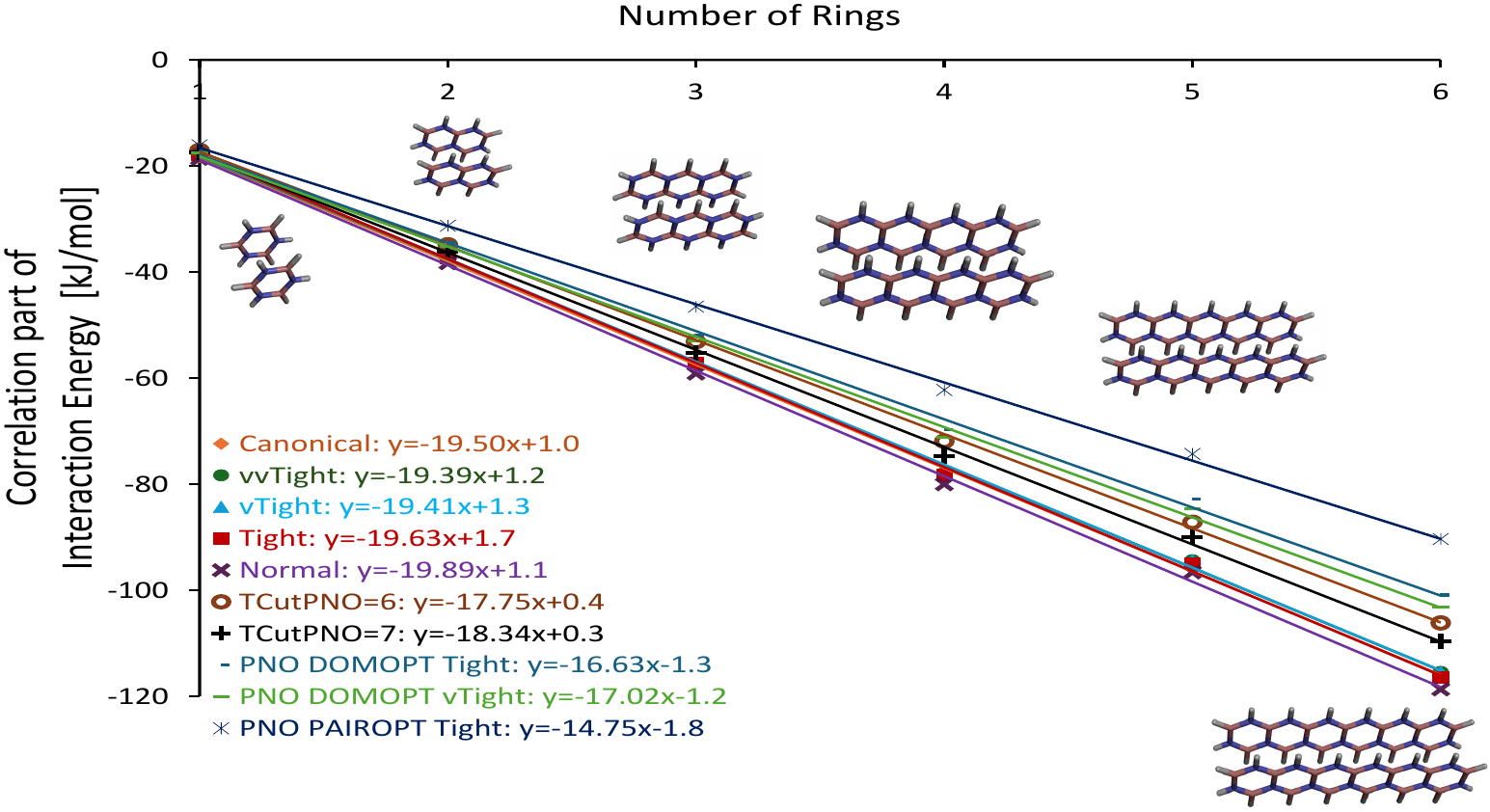}
  \caption{cp-uncorr DZ energies for BNx $\cdots$ NBx}
\end{subfigure}

\medskip

\begin{subfigure}[t]{\linewidth}
  \centering\includegraphics[width=\linewidth]{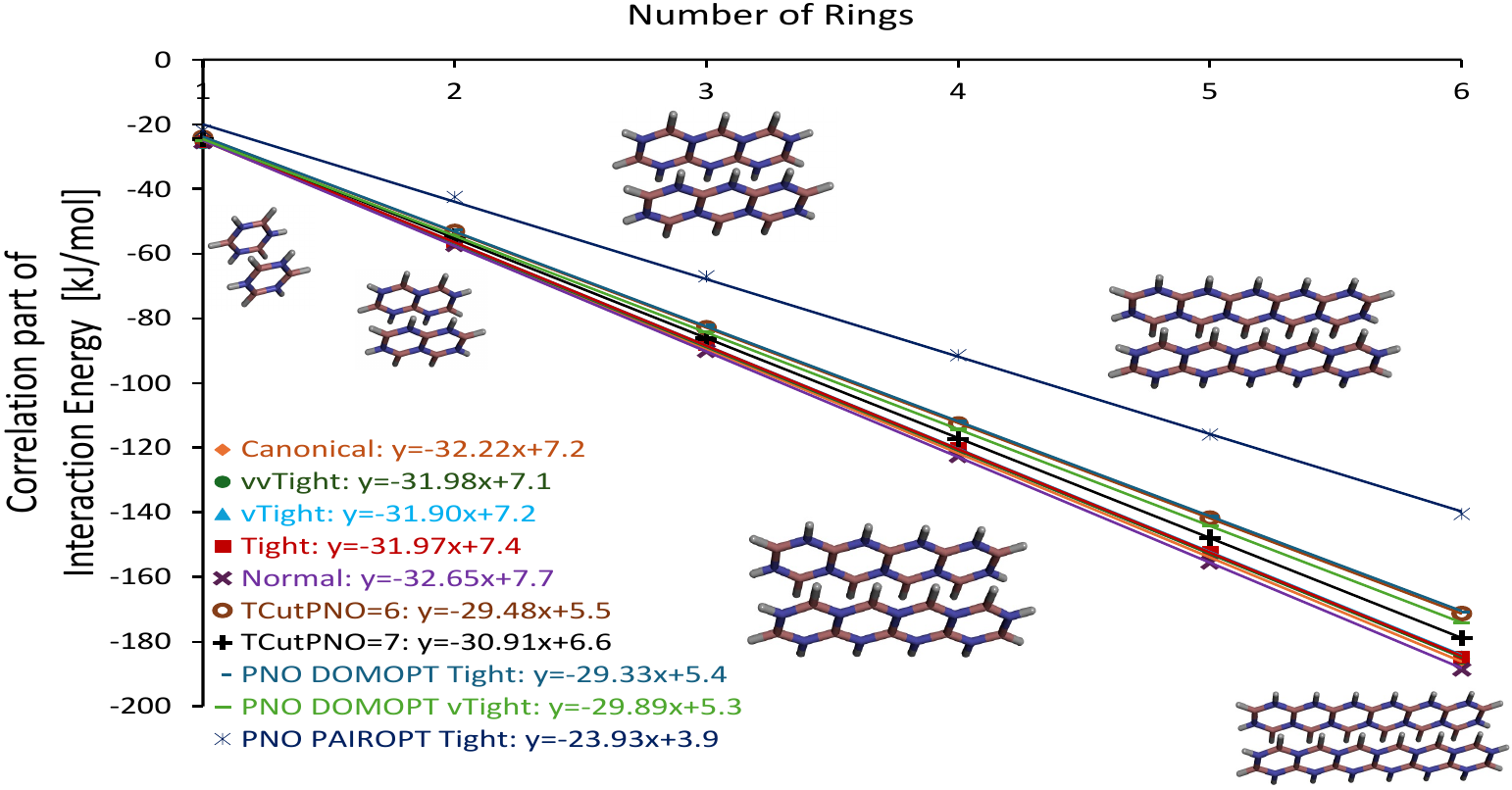}
  \caption{cp-uncorr DZ energies for BNx $\cdots$ BNx}
\end{subfigure}
\end{minipage}

\caption{Correlation part of interaction energies (\kjmol) derived from various coupled cluster methods in as a function of system size per number of rings.}

\label{fgr:LocalCorr}

\end{figure}

Having established the basis set sensitivity, we turn to estimating canonical CCSD(T) results at the complete basis set limit. With the advent of local correlation methods in Coupled-Cluster theory and extrapolating their cutoff parameters, calculations of large systems become feasible and we can compare local methods to canonical CCSD(T) results. For this purpose, we approximate canonical CCSD(T) by several local coupled-cluster methods with the small DZ basis sets (both cp-corrected and cp-uncorrected DZ, see Figure \ref{fgr:LocalCorr}), where we are able to compute canonical CCSD(T) for all systems. Again, perfectly linear relationships are observed for all variants of CCSD(T).

The first local variant investigated is local natural orbital (LNO) coupled-cluster, where the LNO cutoff threshold setting increases as Normal $\rightarrow$ Tight $\rightarrow$ vTight $\rightarrow$ vvTight. Here, the slope converges toward to canonical CCSD(T) most of the times. When a cp-corrected DZ basis set is applied, the successive differences in slopes for polyacene PD systems when going from Normal to Tight to vTight to vvTight LNO thresholds to canonical CCSD(T) are 2.73, 0.89, 0.14, 0.16 \kjmol\ per subunit, respectively, showing a systematic convergence. For the BNx $\cdots$ BNx series, no systematic convergence in the slope differences is observed, as the slopes change by 2.28, 0.97, -0.10, -0.04 \kjmol\ per subunit. Despite the fact that, as expected, LNO-CCSD(T)-vvTight gives the closest result to canonical-CCSD(T), the slope differences for different CCSD(T) methods is somewhat unexpectedly increase along Normal $\rightarrow$ Tight $\rightarrow$ canonical $\rightarrow$ vvTight $\rightarrow$ vTight.
For the BNx $\cdots$ NBx series, the differences in slopes are 2.16, 0.63, -0.01, 0.10 \kjmol\ per subunit, implying that the LNO-CCSD(T)-vTight slope value is closer to canonical-CCSD(T) than LNO-CCSD(T)-vvTight. 

    It should be pointed out that in our previous paper \cite{fishman2025another}, the last vTight point at the cp-corrected aTZ value was unfortunately incorrect, resulting in an erroneous regression line in Figure 2c for LNO-CCSD(T)-vTight. The corresponding LNO-CCSD(T)-vTight line hence changes from $ y = 	-22.05 \times x - 1.60 \ \kjmol $ to $ y = -23.78 \times x + 2.4 \ \kjmol $. The corrected Figure 2c for Ref. \citenum{fishman2025another} is provided in the Supporting Information of the present paper.

Based on the foregoing analysis, LNO-CCSD(T) \textbf{underestimates} canonical-CCSD(T) in the calculation of non-covalent interaction energy for the polyacene PD and BNx $\cdots$ NBx series (as well as for sandwich-structured polyacenes, as described in Ref. \citenum{fishman2025another}). For the more electrostatically bound BNx $\cdots$ BNx series, LNO \textbf{overestimates} canonical-CCSD(T).

Using the same DZ cp-corrected basis set for the second tested local method, Pair Natural Orbital Local CCSD(T) [PNO-LCCSD(T)], when going from the Pairopt-Tight threshold to Domopt-Tight to Domopt-vTight to canonical CCSD(T), the slopes change by -0.70, -0.34, -0.73 \kjmol \ per subunit for the polyacene PD series, respectively. Thus, PNO-LCCSD(T) is showing a systematic convergence, but with negative sign, unlike the LNO approximation. For the PNO approximation, a similar trend of convergence is observed for the BNx $\cdots$ NBx series, where the slopes change by -0.29, -0.41, -0.01 \kjmol \ per subunit. In contrast, for the BNx $\cdots$ BNx series, a non-systematic convergence of slopes is observed, changing by -1.09, -0.56, +0.11 \kjmol \ per subunit. For all types of local methods applied, the PNO-LCCSD(T) approximation gives the closest slope to canonical CCSD(T) with the Domopt-vTight parameter.

As a third method, Domain-Based Local Pair Natural Orbital Coupled-Cluster [DLPNO-CCSD(T$_1$)] has been utilized. Here -- even when employing counterpoise corrections -- we cannot make a definitive conclusion regarding the optimal TCutPNO parameter for the PNO space of the DLPNO approximation. For the polyacene PD series, the TCutPNO parameter of $10^{-6}$ yields a slope closer in value to canonical CCSD(T) than a parameter of $10^{-7}$. For the BNx $\cdots$ BNx series, $10^{-7}$ yields closer result, while for BNx $\cdots$ NBx series both cutoff parameters are equally far from the canonical CCSD(T) slope.

As discussed in the previous section, the deviation of counterpoise (cp)-corrected TZ slopes is not as far from CBS as DZ ones. Moreover, the Natural Orbital Approximations compress DZ less than the larger basis sets and thus DZ basis might not be fully representative. Hence, the analysis of results obtained with the TZ basis is giving a more complete picture, although we have fewer species to compare to.  As far as canonical-CCSD(T)/TZ calculations are feasible for at least four systems in each series, we can consider the convergence for different local coupled-cluster theories to canonical-CCSD(T) for the 1$\rightarrow$4 TZ Slope. See relevant Figure S1 on page 36 in SI. The convergence of the slope in the PNO-LCCSD(T) approximation has the trend Pairopt-Tight $\rightarrow$ Domopt-Tight $\rightarrow$ Domopt-VTight $\rightarrow$ canonical for all systems.

For LNO-CCSD(T) in combination with the cp-corrected TZ basis set, the thresholds converge as expected with increasing cutoffs: Normal $\rightarrow$ Tight $\rightarrow$ vTight $\rightarrow$ vvTight $\rightarrow$ canonical. The same is happening for the cp-uncorrected TZ basis set with the exception of the BNx $\cdots$ BNx series, in which, interestingly, the Tight threshold reproduces the canonical result with the lowest deviation.

For the DLPNO approximation, the BNx $\cdots$ NBx series is not as convergent for the different cut-offs in case of cp-corrected TZ basis set: The TCutPNO parameter of $10^{-6}$ overestimates canonical results, while the TCutPNO parameter of $10^{-7}$ underestimates it, but for larger systems (with 3 or 4 rings) TCutPNO=$10^{-7}$ results are reliable. This behavior of DLPNO convergence may be affected not only by TCutPNO parameter, but also by TCutDO, that dominantly controls the length of the interaction and can lead to different convergence directions: tightening TCutDO converges the DLPNO energy toward the canonical limit from below, whereas tightening other thresholds can converge from above.\cite{saitow2017new, gray2024assessing}  
In all other cases, each series converges from $10^{-6}$ $\rightarrow$ $10^{-7}$ $\rightarrow$ canonical.

Related to this, LNO-CCSD(T) (vTight and vvTight threshold settings) and PNO-LCCSD(T) (DOMOPT=vTight setting) are both equally reliable and have provided the similar deviation from canonical CCSD(T). However, in our further analysis, we focus on LNO approximation as far as it has more predictable convergence threshold-to-threshold error behavior.\cite{semidalas2022mobh35} 


\subsection{Electronic Structure Methods}
Having estimated the slopes of canonical basis set limit CCSD(T) and thus its performance for large molecules, we evaluate the efficiency of more approximate electronic structure methods.
Here, we investigate several post-Hartree-Fock methods, such as MP2, MP2.5, MP3; the dispersion contribution of DFT-SAPT, as well as dRPA. To determine the degree of agreement of the approximate methods with the `gold standard' canonical CCSD(T), we compare absolute values of slopes obtained using these approximate methods, ultimately determining their quality within this metric.
These approximate methods can help us to understand which physical aspects -- second-order dispersion, third-order perturbative corrections, connected doubles, perturbative triples, dynamical screening, electrostatics, induction, and dispersion -- are actually responsible for the interaction trend.

Slope values for the parallel displaced polyacene systems exhibit the trend MP3 $<$ CCSD $<$ dRPA $\approx$ CCSD(T) $\approx$ DFT-SAPT $<$ MP2.5 $<$ MP2 for cp-corrected and cp-uncorrected (excluding DFT-SAPT for the latter, as DFT-SAPT is only defined when using cp-corrections). As was described earlier for sandwich polyacene dimers (see section "Electronic Structure Methods" in Ref. \citenum{fishman2025another}), for the parallel displaced polyacene dimers MP2 also significantly overestimates interaction energies whereas MP3 significantly underestimates them and, which is as well important to note, DFT-SAPT and dRPA are closer to CCSD(T) than the post-Hartree-Fock methods investigated. The polyacene series has strongly size-dependent post-MP2 regime in which finite order M{\o}ller–Plesset perturbation theory is unreliable.

An identical trend for cp-corrected slope values is observed for the BNx $\cdots$ NBx dimers. Here, deviations of MP2 and MP3 slopes are not as large from CCSD(T) (12\% and 23\% respectively), while for polyacene dimers the same deviations are as big as 50\%. BNx $\cdots$ NBx still has non-negligible post-MP2 correlation, but the positive CCSD/MP3-type correction is partially cancelled by the negative triples contribution. The relatively small final MP2 error is thus probably due to error cancellation effects. For BNx $\cdots$ NBx, dRPA demonstrates the best approximation to CCSD(T). dRPA reproducing CCSD(T) indicates that screened correlation/ring correlation is sufficient to describe the total size-dependent trend. In the case of cp-uncorrected slope values for the BNx $\cdots$ NBx dimers, the MP2 and MP2.5 methods are the best approximation to CCSD(T), while MP3 still underestimates interaction energies. 

In contrast to the polyacene PD series, the slopes of the BNx $\cdots$ BNx systems exhibit a very different behavior: MP3 $<$ CCSD $<$ MP2.5 $<$ CCSD(T) $<$ MP2 $<$ dRPA $<$ DFT-SAPT. Here, DFT-SAPT and dRPA are, interestingly, inferior to the MP2.5 and MP2 methods, whereas MP3 still noticeably underestimates interaction energies. The apparent success of MP2 or MP2.5 in BNx $\cdots$ BNx series is most likely due to favorable error cancellation as far as MP2-MP3 difference is large and systematic, not because the underlying physics is purely second-order.

The larger post-MP2 sensitivity observed for polyacenes is consistent with its substantially smaller HOMO–LUMO gap (See SI on pages 86-87) that indicates lower-energy electronic excitations and enhanced polarizability. Since the MP2 correlation energy depends inversely on occupied–virtual energy denominators, smaller gaps can amplify second-order amplitudes and make higher-order screening and triples effects essential.


\begin{figure}[htbp]
\centering

\begin{minipage}[t]{0.49\textwidth}
\centering
\begin{subfigure}[t]{\linewidth}
  \centering\includegraphics[width=\linewidth]{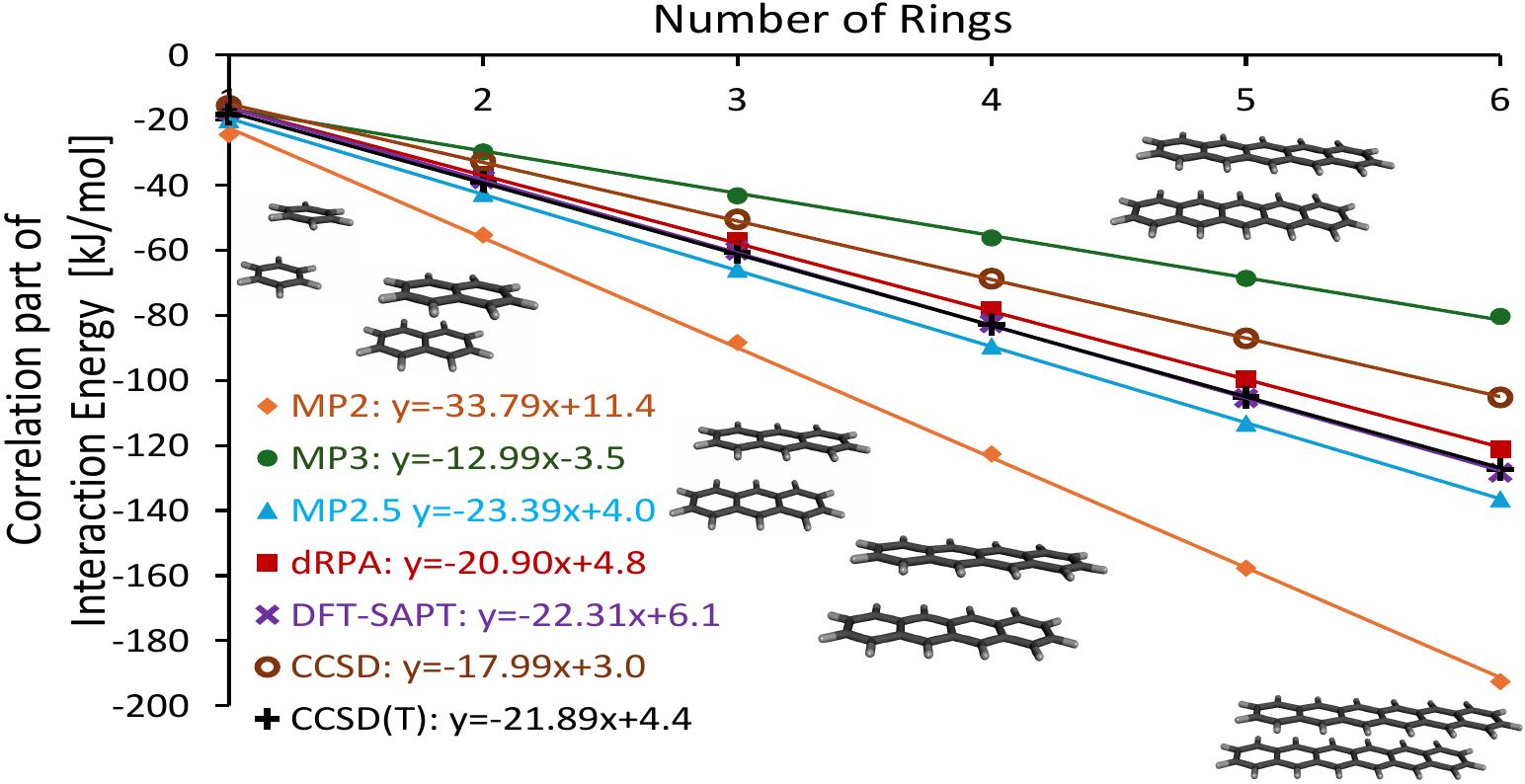}
  \caption{cp-corr DZ energies for acenes PD}
\end{subfigure}

\medskip

\begin{subfigure}[t]{\linewidth}
  \centering\includegraphics[width=\linewidth]{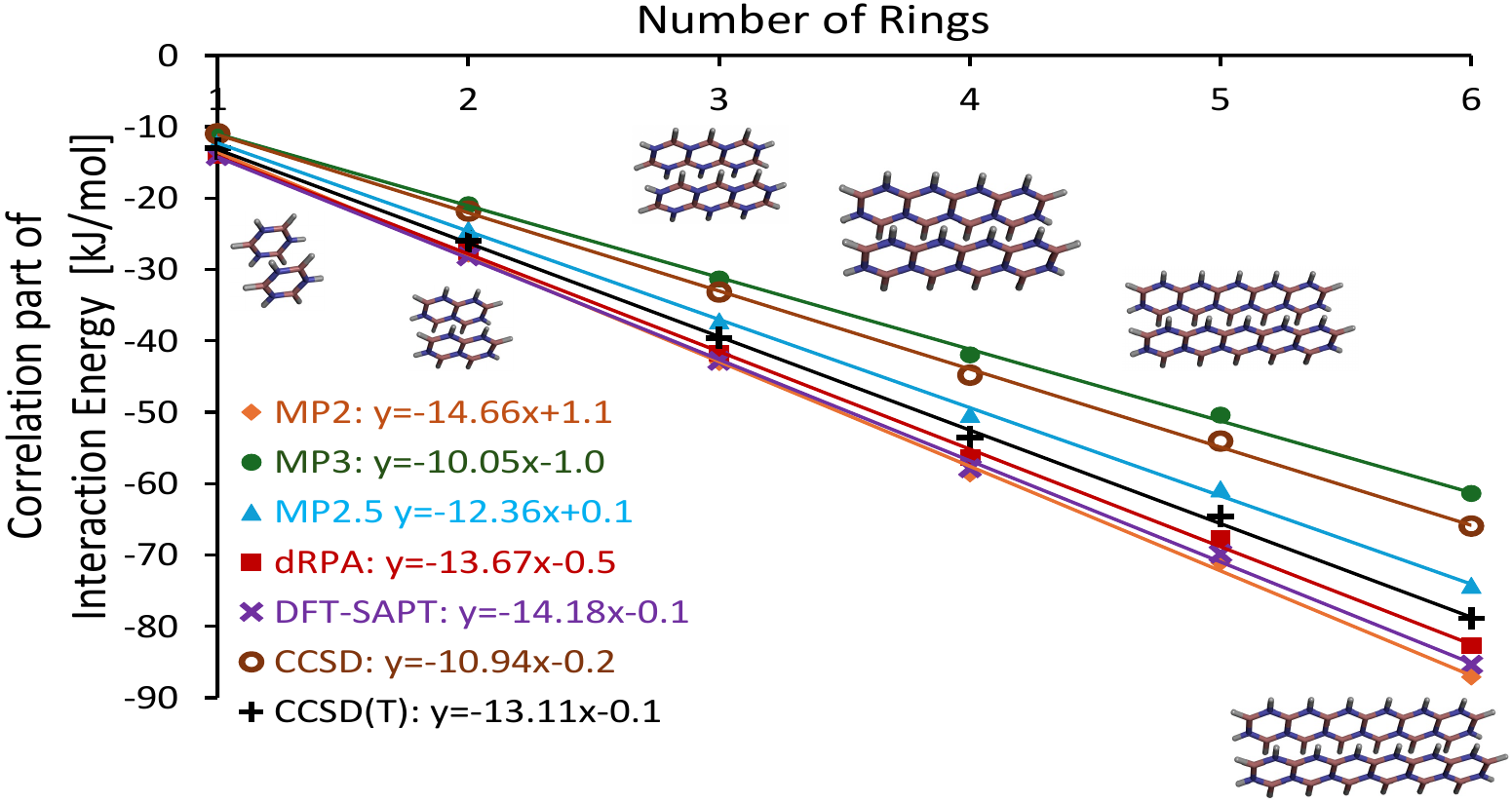}
  \caption{cp-corr DZ energies for BNx $\cdots$ NBx}
\end{subfigure}

\medskip

\begin{subfigure}[t]{\linewidth}
  \centering\includegraphics[width=\linewidth]{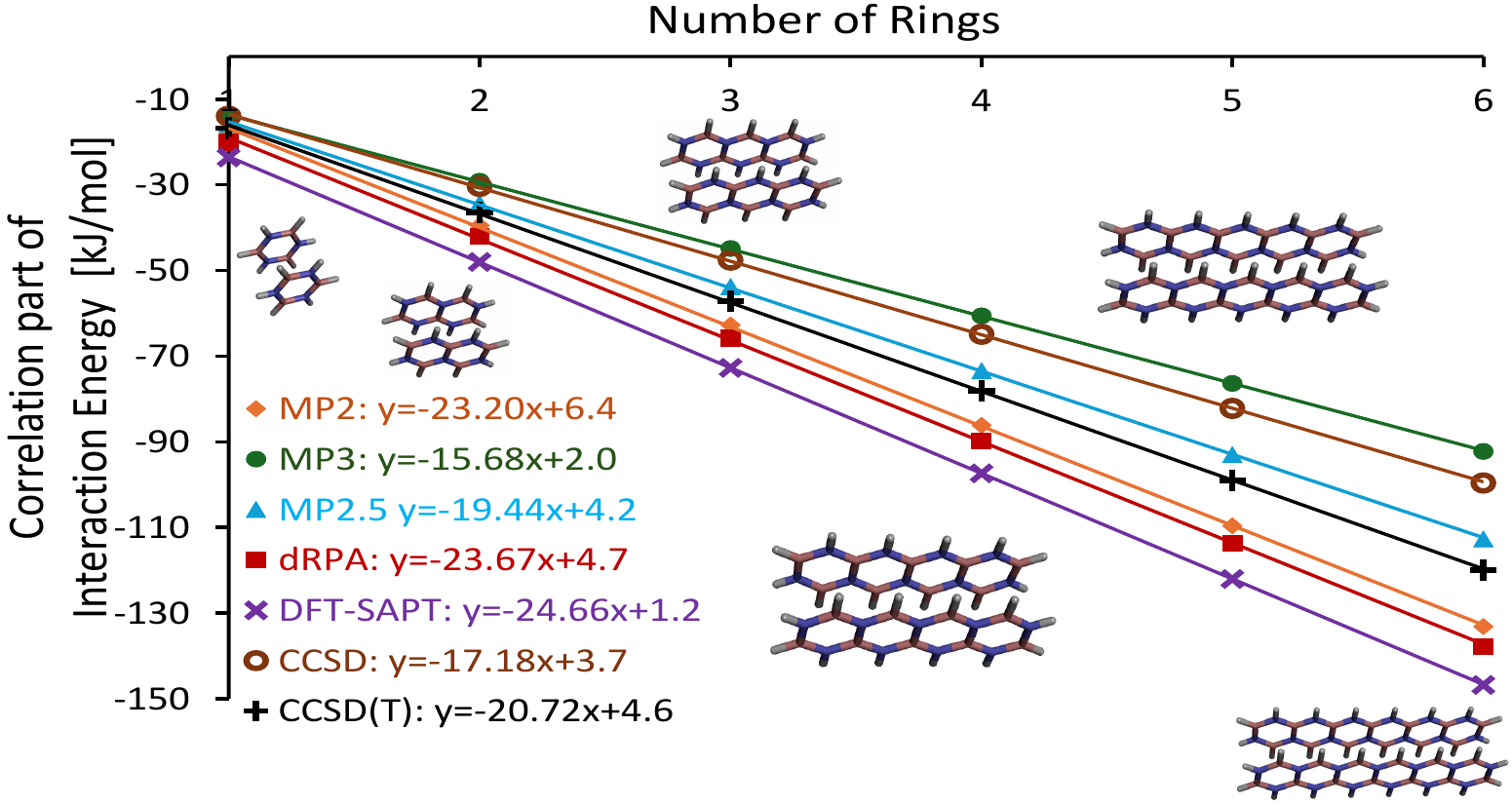}
  \caption{cp-corr DZ energies for BNx $\cdots$ BNx}
\end{subfigure}
\end{minipage}
\hfill
\begin{minipage}[t]{0.49\textwidth}
\centering
\begin{subfigure}[t]{\linewidth}
  \centering\includegraphics[width=\linewidth]{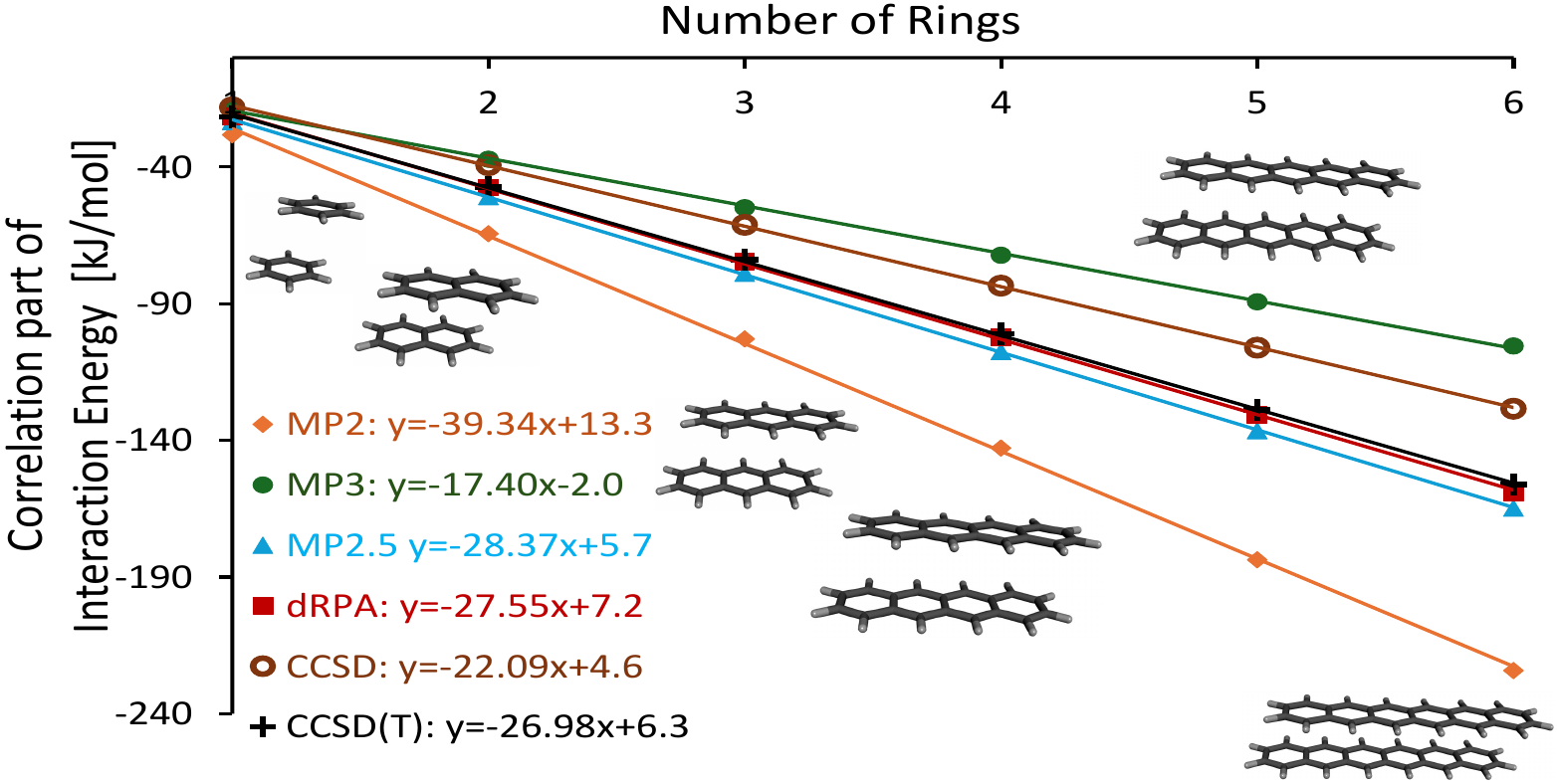}
  \caption{cp-uncorr DZ energies for acenes PD}
\end{subfigure}

\medskip

\begin{subfigure}[t]{\linewidth}
  \centering\includegraphics[width=\linewidth]{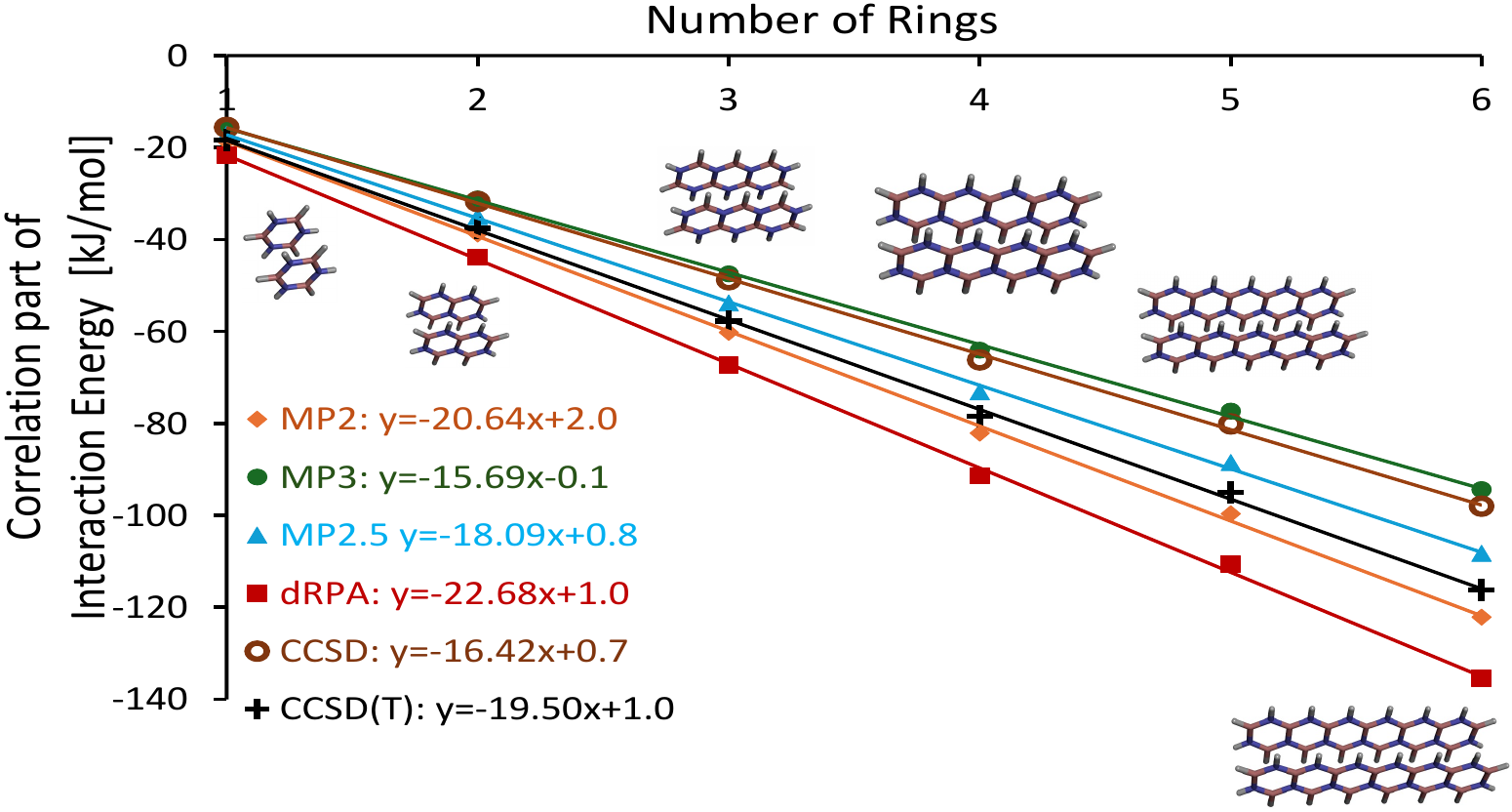}
  \caption{cp-uncorr DZ energies for BNx $\cdots$ NBx}
\end{subfigure}

\medskip

\begin{subfigure}[t]{\linewidth}
  \centering\includegraphics[width=\linewidth]{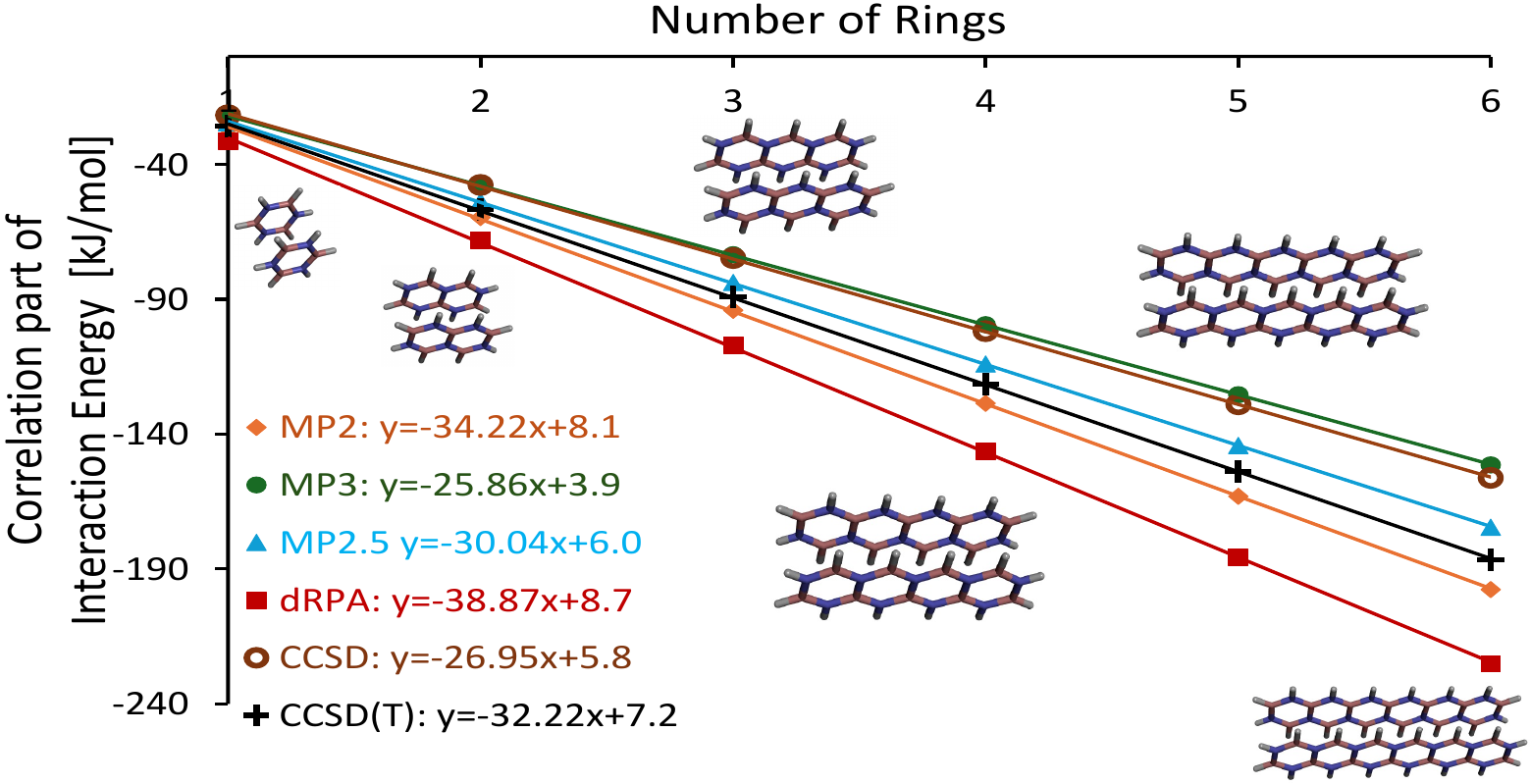}
  \caption{cp-uncorr DZ energies for BNx $\cdots$ BNx}
\end{subfigure}
\end{minipage}

\caption{Correlation part of interaction energies derived from various post-Hartree-Fock (and DFT) methods in \kjmol\ vs. number of rings.}

\label{fgr:ESM}

\end{figure}

\subsection{Best Estimates of Slopes}

Unfortunately, many calculations using the ``gold-standard'' canonical CCSD(T) method are doable only with the DZ basis set. Thus, we can either calculate each system which belongs to the investigated sequence with this (or another) rather small basis set, or increase the basis set size and calculate a lesser number of systems, thus adding an additional dimension to the problem. Unfortunately, CCSDT and CCSDT(Q) calculations for the parallel-displaced naphthalene dimer or diborazine-fused sandwich dimers are currently computationally not feasible even with a ``truncated'' DZ basis set with no $p$-functions on Hydrogen, Thus, we consider the post-CCSD(T) methods CCSDT-2 and CCSDT-3 \cite{noga1987towards} instead of CCSDT, as they scale as $O(N^7)$ rather than the $O(N^8)$ with the number of electrons. Deviations of CCSDT(Q) from CCSD(T) can be estimated using CCSD(T)$_\lambda$, as the correlation part of CCSD(T)$_\lambda$ seems to be somewhat closer to the correlation part of CCSDT(Q) than CCSD(T) (See Figure 4, SI of Ref. \citenum{fishman2025another} and Ref. \citenum{jmlm336}).

An anomaly is seen for the MP2 correlation energy of coronene dimer when diffuse functions are added to the basis set. Ordinarily, it is well known that adding diffuse functions accelerates basis set convergence in covalent bond energies and noncovalent interactions alike.\cite{boese20152} However, for this extended planar system, we observe the opposite: in particular, the `spread’ between cp-corrected and cp-uncorrected calculations is much larger for aug-cc-pV\{$n-1$,$n$\}Z extrapolations than for their cc-pV\{$n-1$,$n$\}Z counterparts. The root cause is numerical rather than quantum mechanical: diffuse functions cause serious near-linear dependence in the basis set, as reflected in the smallest eigenvalues of the overlap matrix $S$ dipping into the $10^{-11}-10^{-12}$ region for aug-cc-pV$n$Z (vs. $10^{-6}-10^{-8}$ for cc-pV$n$Z). As a result, the SCF in the former case becomes numerically ill-conditioned: the remedy of codes likes MRCC is to discard all eigenvectors of $S$ below a certain cutoff (by default $10^{-7}$). Alas, in this case there are many more eigenvectors to be discarded in the dimer than the two sets of monomer ‘discards’, which causes a size-consistency issue in the cp-uncorrected calculation. (The cp-corrected is not affected in the same manner, as $S$ is the same for dimer and monomer+ghost).

Thus, in this study, we use the cc-pV\{Q,5\}Z basis set extrapolation instead of aug-cc-pV\{Q,5\}Z one. The slopes of the correlation energy, extrapolated to cp-uncorrected and cp-corrected complete basis set limits, for all investigated series are summarized in Table \ref{tbl:slope-all-systems}, while Table \ref{tbl:slope-all-systems-inclHF} summarizes the slopes of the full interaction energies obtained by different WFT-based methods. 

\begin{table}[!htbp]
\centering
\caption{Best available cc-pV\{Q,5\}Z Energy Slopes Correlation Energy Slopes (kJ/mol per Subunit). All numbers are estimates, except LNO-CCSD(T)-Tight.
The top part of the Table shows the cp-corrected results, the bottom part the cp-uncorrected ones. Detailed Tables are given in the SI on pages 61-66.}
\label{tbl:slope-all-systems} 

\resizebox{\textwidth}{!}{%
\begin{tabular}{rrrrrrrrrrr}
\hline\hline

 & \multicolumn{10}{c}{Method} \\  \cline{2-11}
 & & & \multicolumn{3}{c}{CCSD(T)} & & \multicolumn{3}{c}{post-CCSD(T)} \\ \cmidrule(lr){4-6} \cmidrule(lr){8-11}

System & MP2 & CCSD & \texttt{Tight} & \texttt{vTight} & canonical & CCSD(T)$_\lambda$$^a$ & CCSDT-2$^a$ & CCSDT-3$^a$ & CCSDT$^{a,b}$ & CCSDT(Q)$^{a,b}$ \\   \hline

BNx $\cdots$ NBx & -18.90 & -13.64 & -17.53 & -17.90 & -17.66 & -17.58 & -17.29 & -17.42  & & \\
BNx $\cdots$ BNx & -30.83 & -22.86 & -28.20 & -27.98 & -28.26 & -28.02 & -27.51 & -27.71  & & \\
Acene PD & -45.40 & -24.86 & -32.45 & -31.75 & -31.51 & -31.21 & -30.71 & -30.77 & & \\
Acenes$^c$ & -34.22 & -18.41 & -24.89 & -24.43 & -23.13 & -22.90 & -22.44 & -22.47 & -22.08 & -22.64 \\
Polyene Relax. & -11.02 & -7.06 &  &  & -8.75 & -8.64 & -8.55 & -8.57 & -8.55 & -8.78 \\
Polyene Fixed & -4.87 & -3.06 &  &  & -3.82 & -3.78 & -3.73 & -3.74 & -3.72 & -3.82 \\
Coronene PD & -31.96 & -16.59 & -22.27 & -21.48 & -20.97 & -20.78 &  &  & -20.25 & -20.76 \\
Coronene Sandwich &  -28.52 & -15.10 & -20.46 & -19.72 & -19.04 & -18.90 & & & & \\

\hline

BNx $\cdots$ NBx & -19.35 & -14.42 & -17.81 & -17.49 & -17.79 & -17.66 & -17.35 & -17.48 &  & \\
BNx $\cdots$ BNx & -30.64 & -22.63 & -27.98 & -27.67 & -27.95 & -27.64 & -27.05 & -27.25  & & \\
Acene PD & -45.34 & -24.49 & -32.44 & -31.57 & -31.09 & -30.63 & -29.98 & -30.05  & & \\
Acenes$^c$ & -34.96 & -19.16 & -24.86 & -24.44 & -23.98 & -23.71 & -23.21 & -23.25 & -22.88 & -23.44 \\
Polyene Relax. & -11.04 & -7.00 &  &  & -8.68 & -8.57 & -8.46 & -8.49 & -8.45 & -8.69 \\
Polyene Fixed & -4.86 & -3.01 &  &  & -3.83 & -3.88 & -3.83 & -3.84 & -3.76 & -3.94 \\
Coronene PD & -32.15 & -16.76 & -22.29 & -21.59 & -21.01 & -20.74 &  &  & -19.95 & -20.58 \\
Coronene Sandwich & -28.66 & -15.10 & -20.43 & -19.68 & -18.87 & -18.69 & & & & \\

\hline\hline

\end{tabular}
}
$^a$ The post-CCSD(T) and CCSD(T)$_\lambda$ slopes have been scaled by a factor of 1.4 to account for the small cc-pVDZ basis set size --- see discussion in reference \citenum{fishman2025another} and in its Supporting Information on pages 83-89.\\
$^b$ CCSDT and CCSDT(Q) slopes for Coronene PD have been estimated according to respective difference with CCSD(T) for hexacene sandwich-structured dimer - see discussion below.
\\
$^c$ Re-estimated values for sandwich-structured acene series with the most recent MRCC version. \\

\end{table}

\begin{figure}[p]

     \centering

    \begin{subfigure}[a]{\linewidth}
        \centering
         \includegraphics[width=1.0\textwidth]{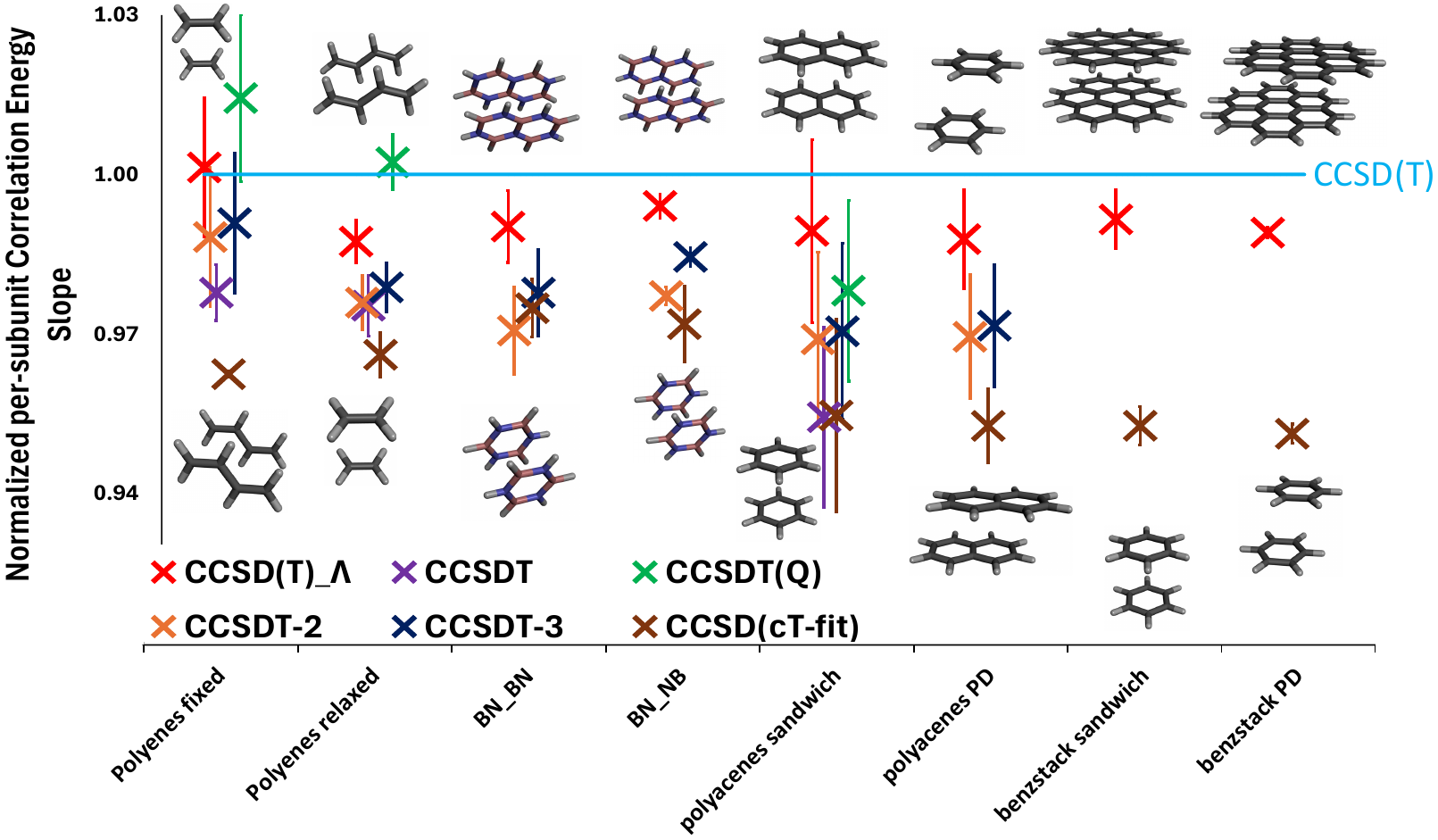}
    \end{subfigure}
  
    \caption{Deviations of Correlation energies from reference CCSD(T) method in $\%$ vs. number of subunits. The first two dimers (e.g. benzene and naphthalene dimer) of each series are displayed above and below each series column to illustrate which series is displayed.
    }

    \label{fgr:graphTOC}
\end{figure}

The CCSD(T) CBS lines for all our systems up to the hexacene analogues (indicated by 1 $\rightarrow$ 6 ) are obtained as follows:
\[y = (-7.0 \pm 0.1 \ \kjmol) \times x -0.4 \pm 0.4 \ \kjmol\]  per ring including HF for BNx $\cdots$ NBx series,

\[y = (-16.8 \pm 0.4 \ \kjmol)\times x + 3.4 \pm 0.4 \ \kjmol\]  per ring including HF for BNx $\cdots$ BNx series, and

\[y = (-16.7 \pm 0.3 \ \kjmol)\times x + 6.2 \pm 0.1 \ \kjmol\]  per ring including HF for the parallel displaced polyacene series. 

The best estimate CCSD(T)$_\lambda$/cc-pV\{Q,5\}Z interaction energies for the benzene and naphthalene parallel displaced dimers are -11.4 $\pm$ 0.3 and -26.3 $\pm$ 0.8 \kjmol\ respectively, resulting in a slope value of -14.9 \kjmol\ per subunit, similar to the one discussed above. 

It is possible to arrange the systems according to the strength of correlation: polyacenes pd $>$ BN$\cdots$BN $>$ polyacenes sandwich $>$ coronene pd $>$ coronene sandwich  $>$ BN$\cdots$NB $>$ polyene relaxed $>$ polyene fixed. The parallel displaced acene series has the strongest corrrelation, while polyenes are weakly bounded. Thus, slope value (\kjmol\ per subunit) may be considered as correlation indicator in any correlated-consistent basis set. 


\begin{table}[!htbp]
\centering
\caption{Best available cc-pV\{Q,5\}Z Energy Slopes, this time including Hartree-Fock (kJ/mol per Subunit), similar to Table \ref{tbl:slope-all-systems}. All numbers are estimates, except LNO-CCSD(T)-Tight. The top part of the Table shows the cp-corrected results, the bottom part the cp-uncorrected ones. }
\label{tbl:slope-all-systems-inclHF} 

\resizebox{\textwidth}{!}{%
\begin{tabular}{rrrrrrrrrrr}
\hline\hline

 & \multicolumn{10}{c}{Method} \\  \cline{2-11}
 & & & \multicolumn{3}{c}{CCSD(T)} & & \multicolumn{3}{c}{post-CCSD(T)} \\ \cmidrule(lr){4-6} \cmidrule(lr){8-11}

System & MP2 & CCSD & \texttt{Tight} & \texttt{VeryTight} & canonical & CCSD(T)$_\lambda$$^a$ & CCSDT-2$^a$ & CCSDT-3$^a$ & CCSDT$^{a,b}$ & CCSDT(Q)$^{a,b}$ \\   \hline

BNx $\cdots$ NBx & -8.36 & -3.10 & -6.99 & -7.36 & -7.11 & -7.03 & -6.75 & -6.88 &  & \\
BNx $\cdots$ BNx & -19.84 & -11.87 & -17.20 & -16.99 & -17.26 & -17.03 & -16.51 & -16.72  & & \\
Acene PD & -30.97 & -10.44 & -18.02 & -17.33 & -17.08 & -16.78 & -16.28 & -16.34  & &  \\
Acenes$^c$ & -22.27 & -6.47 & -12.95 & -12.48 & -11.18 & -10.96 & -10.49 & -10.53 & -10.14 & -10.70 \\
Polyene Relax. & -5.72 & -1.76 &  &  & -3.45 & -3.35 & -3.25 & -3.28 & -3.25 & -3.48 \\
Polyene Fixed & -3.38 & -1.56 &  &  & -2.33 & -2.28 & -2.23 & -2.25 & -2.23 & -2.33 \\
Coronene PD & -23.54 & -8.17 & -13.85 & -13.06 & -12.55 & -12.36 &  &  & -11.83 & -12.34 \\
Coronene Sandwich & -18.87 & -5.45 & -10.81 & -10.07 & -9.39 & -9.25 &  &  &  &  \\

\hline

BNx $\cdots$ NBx & -8.49 & -3.56 & -6.95 & -6.63 & -6.93 & -6.80 & -6.49 & -6.62 & & \\
BNx $\cdots$ BNx & -19.06 & -11.04 & -16.40 & -16.08 & -16.37 & -16.06 & -15.47 & -15.67  & & \\
Acene PD & -30.64 & -9.79 & -17.74 & -16.86 & -16.39 & -15.93 & -15.27 & -15.35  & & \\
Acenes$^c$ & -23.02 & -7.22 & -12.92 & -12.50 & -12.04 & -11.77 & -11.27 & -11.32 & -10.94 & -11.50 \\
Polyene Relax. & -5.63 & -1.59 &  &  & -3.27 & -3.16 & -3.05 & -3.08 & -3.04 & -3.28 \\
Polyene Fixed & -3.29 & -1.45 &  &  & -2.26 & -2.32 & -2.27 & -2.28 & -2.20 & -2.38  \\
Coronene PD & -23.55 & -8.16 & -13.69 & -12.99 & -12.41 & -12.14 &  &  & -11.35 & -11.98 \\
Coronene Sandwich & -18.86 & -5.30 & -10.63 & -9.87 & -9.07 & -8.88 & &  &  &  \\

\hline\hline

\end{tabular}
}
$^a$ The post-CCSD(T) and CCSD(T)$_\lambda$ slopes have been scaled by a factor of 1.4 to account for the small cc-pVDZ basis set size --- see discussion in reference \citenum{fishman2025another} and in its Supporting Information on pages 83-89. 
\\
$^b$ CCSDT and CCSDT(Q) slopes for Coronene PD have been estimated according to respective difference with CCSD(T) for hexacene sandwich-structured dimer.
\\
$^c$ Re-estimated values for sandwich-structured acene series with the most recent MRCC version. \\

\end{table}

\subsection{Coronene Dimer}
Of special interest is the radially fused polycyclic aromatic hydrocarbon series going from Benzene $\rightarrow$ Coronene $\rightarrow$ Circumcoronene, as the coronene dimer was heavily investigated in the past.

\begin{figure}[p]

     \centering

    \begin{subfigure}[a]{\linewidth}
        \centering
         \includegraphics[width=0.6\textwidth]{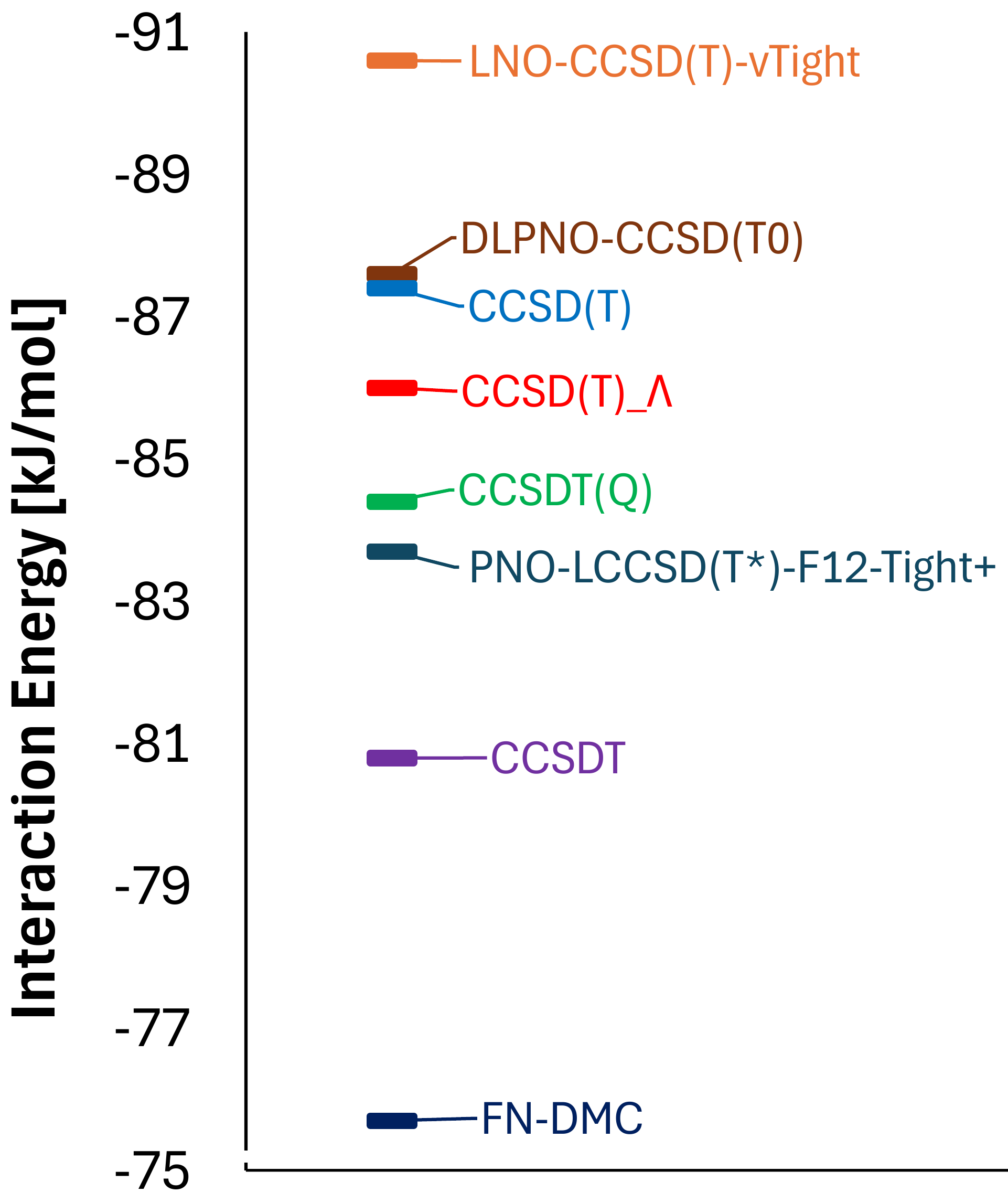}
    \end{subfigure}
  
    \caption{Interaction energies in \kjmol\ for Coronene Parallel Displaced dimer at the different levels of theory. The CCSDT and CCSDT(Q) values are estimates from the similar behaving acene sandwich interaction energies.}

    \label{fgr:c2c2pd-res}
    
\end{figure}

Following the same recipe of Slope and Intercept estimation as above (see also SI on pages 76-78), we are able to estimate the interaction energy of the parallel displaced coronene dimer.
Table \ref{tab:c2c2pd-results} and Figure \ref{fgr:c2c2pd-res} show the interaction energy of the parallel displaced coronene dimer at different levels of computational theory with various estimation techniques. Here, our CCSD(T)/CBS result is close to the results reported by previous groups, with LNO-CCSD(T) overestimating the canonical interaction energy for hydrocarbons.

During the re-evaluation of the interaction energies for sandwich-structured polyacene series, we noticed that the [(T)$_\lambda$-(T)]/CBS(cp-corr) difference for sandwich-structured hexacene dimer, which is 1.34 kJ/mol, is surprisingly close to the [(T)$_\lambda$-(T)]/CBS(cp-corr) difference for the parallel displaced coronene dimer of 1.37 kJ/mol. Assuming a similar behavior of other post-CCSD(T) corrections for $\pi$-conjugated aromatic systems, we estimate the (Q)-(T) difference for coronene PD to be expressed mathematically through a (Q)-(T) difference for sandwich-structured hexacene, such as: 

\begin{equation}
   [(Q)-(T)]_{\mathrm{coronene \ PD}} \approx [(T)_\lambda-(T)]_{\mathrm{coronene \ PD}} \times [(Q)-(T)]_{\mathrm{hexacene}}  / [(T)_\lambda-(T)]_{\mathrm{hexacene}}
\end{equation}

This results in an estimate of $[(Q)-(T)]_{\mathrm{coronene \ PD}} \approx 1.37 \times 2.87 / 1.34 = 2.95 \ \kjmol$. Thus, the CCSDT(Q)/CBS(cp-corr) interaction energy for the parallel displaced coronene dimer is estimated as -84.4 \kjmol. The same estimation for the $T_3 - (T)$ contribution yields an interaction energy of -80.8 \kjmol\ at the CCSDT/CBS(cp-corr) level. The same formula for counter-poise uncorrected CCSDT and CCSDT(Q) interaction energies at the complete basis set provides -77.5 and -81.8 \kjmol. 

\begin{table*}[!htbp]
\caption{Interaction energies for the coronene parallel displaced dimer at different levels of theory in the complete basis set limit. All values are reported in \kjmol.}
\centering

\resizebox{\textwidth}{!}{ 

\begin{tabular}{llll}
\hline
 Level of Theory & Interaction energy  & Reference & Basis\\ \hline
 
MP2 & -161.1 $\pm$ 2.1  & Table 1 in Ref. \citenum{schafer2025understanding} & Plane Wave \\ 
MP2 & -159.3  & Table S1 in Ref. \citenum{villot2022coupled} & CBS \\ 


MP2 & -161.2$\pm$ 0.2 & this work & CBS \{Q,5\} extrap. \\
MP2 & -156.7 $\pm$ 2.5 & this work & CBS a\{Q,5\} extrap. \\
\hline

CCSD & -56.1 $\pm$ 2.1 &  Table 1 in Ref. \citenum{schafer2025understanding} & Plane Wave \\ 
CCSD & -55.4 $\pm$ 0.4 & this work & CBS \{Q,5\} extrap. \\
\hline

DLPNO-CCSD(T$_0$) & -87.6 $\pm$ 1.8 & Table 1 in Ref. \citenum{villot2022coupled} & CBS \\

PNO-LCCSD(T)-F12 (domopt=tight)   &  -83.6 (cp-corr) & Table 10 in Ref. \citenum{ma2018explicitly} & C:aug-cc-pVTZ; H:cc-pVTZ \\ 

PNO-LCCSD(T)-F12 (domopt=tight+)   &  -80.8 (cp-corr) & Table 6 in Ref. \citenum{hansen2025accurate} & C:aug-cc-pV(T+d)Z; H:cc-pVTZ \\ 
PNO-LCCSD(T$^*$)-F12 (domopt=tight+)  &  -83.7 (cp-corr) & Table 6 in Ref. \citenum{hansen2025accurate} & C:aug-cc-pV(T+d)Z; H:cc-pVTZ \\ 

LNO-CCSD(T) (Tight/vTight extrap.)  &  -86.2 $\pm$ 2.5 & Table 1 in Ref. \citenum{AlHamdani2021} & cc-pV\{Q,5\}Z \\ 
LNO-CCSD(T)-vTight  & -90.0 $\pm$ 0.6 & this work & CBS \{Q,5\} extrap. \\
LNO-CCSD(T) (Tight/vTight extrap.)  & -87.5 $\pm$ 0.4 & this work & CBS \{Q,5\} extrap. \\
\hline
CCSD(T) & -86.4 & Table 4,5 in Ref. \citenum{lao2024canonical} & CBS \\
CCSD(T) & -88.3 $\pm$ 2.1 & Table 1 in Ref. \citenum{schafer2025understanding} & Plane Wave \\ 
CCSD(T) & -86.6 $\pm$ 0.8 & this work & CBS \{Q,5\} extrap. \\
CCSD(T)$_\lambda$ & -84.9 $\pm$ 1.1 & this work & CBS \{Q,5\} extrap. \\
CCSDT(Q)$^*$ & -83.1 $\pm$ 1.3 & this work & CBS \{Q,5\} extrap. \\
\hline
\hline
FN-DMC & -75.7 $\pm$ 3.3 & Table 1 in Ref. \citenum{AlHamdani2021}  & \\ 
FN-DMC & -73.4 $\pm$ 0.1 & Table 1 in Ref. \citenum{benali2020quantum}  & def2-QZVP \\ 
CCSD(cT) & -80.8 $\pm$ 2.1 & Table 1 in Ref. \citenum{schafer2025understanding} & Plane Wave \\ 
CCSDT$^*$ & -79.1 $\pm$ 1.7 & this work & CBS \{Q,5\} extrap. \\
\hline
\end{tabular}

} 

Values for this work can be also found in Table S222 on Page 77 of SI.

$^*$ CCSDT and CCSDT(Q) results have been estimated according to respective difference with CCSD(T) for hexacene sandwich-structured dimer.

\label{tab:c2c2pd-results}
\end{table*}

Additionally, we can estimate the interaction energy for the even larger sandwich-stacked circumcoronene ($C_{54}H_{18}$) dimer, where each monomer contains 19 benzene rings.

The CCSD(T) slope and intercept for the sandwich-structured radially fused polycyclic aromatic hydrocarbons is
\[y = 	-9.2 \times x + 2.3 \ \kjmol\]  per ring for cp-corrected interaction energy, resulting in an interaction energy of -172.5 \kjmol\ for the circumcoronene dimer.

If we extrapolate only the CCSD(T) correlation energy for the sandwich-structured coronene series and then add the Hartree-Fock energy of circumcoronene dimer, we obtain $ (-19.0 \pm 0.1 ) \times 19 + (-2.7 \pm 0.2) + (165.7 \pm 1.4) \ = (-197.2 \pm 3.2) \ \kjmol  $ as cp-corrected result. Meanwhile, the LNO-CCSD(T)-Tight correlation energy aligns well with a linear trend (see Figure S5 on page 80 in SI), yielding a coefficient of determination $R^2 > 0.999$. Our best estimate interaction energy for the sandwich-stacked circumcoronene--circumcoronene dimer ath the CCSD(T)$_\lambda$ level of theory is thus -194.0 $\pm$ 3.6 \kjmol\ which is close to the CCSD(T)/CBS result.

Finally, our CCSD(T) is very close to the original LNO-CCSD(T) estimate of Ref. \citenum{AlHamdani2021}, for which the discrepancy with FN-DMC was originally reported. In
 Ref. \citenum{fishman2025another}, we estimated the CCSDT(Q) - CCSD(T) difference which comes from higher-order cluster excitations with a maximum of 3.2 \kjmol. Here, we can confirm this value with a difference of 3.4 \kjmol\ between CCSD(T) and CCSDT(Q), estimating it from the deviation of CCSD(T) to CCSD(T)$_\lambda$. When going from CCSDT(Q) to FN-DMC, we still have 7.4 \kjmol\ left. Here, FN-DMC is probably underestimating the result because of the fixed-node approximation \cite{Zen2025SystematicDiscrepanciesS66,Shee2025NoncovalentAFQMC,Fanta2025NondynamicCorrelationPiPi,LambieAlavi2025AcOHdimer}. This implies that the our best estimate for the coronene dimer holds, as our best estimate is in between canonical CCSD(T) and FN-DMC, but somewhat closer to CCSD(T).

\section{Conclusions}
Extending upon the correlation energy slopes of the alkapolyene (fixed and relaxed geometries) and sandwich-structured polyacene series, we investigated five more series looking at non-covalent interactions. The new series include the minimum, parallel displaced polyacene structures, as well as the two BN analogues of the sandwich polyacenes and both parallel displaced and sandwich coronene series structures. With these, we are able to draw a more complete picture of the nature of non-covalent interactions beyond small model systems, as their behavior when going to larger sizes is explored. All the computed series exhibit a near-linear behavior.

By considering local correlation thresholds, basis sets, and different methods, we can draw conclusions on their scaling behavior. For local orbital approximations  of the coupled-cluster method, only the tightest threshold settings (vvTight for LNO-CCSD(T) and DOMOPT=vTight for PNO-LCCSD(T)) achieve near-canonical-CCSD(T) accuracy. Overall, the basis set limit for the slopes is much quicker accomplished than for the intercepts, making calculations of slopes computationally less demanding.

In general, the polyacene and coronene systems behave quite similarly, even if for the latter, as the system size grows quadratically (specifically, $3x^2 - 3x + 1$) like the number of rings in radially fused polycyclic aromatic hydrocarbons, \emph{viz.}
$1\rightarrow 7\rightarrow 19\rightarrow 37$.

Finally, we are able to show that for the parallel displaced coronene dimer the higher-order triples are antibonding, which bring CCSDT interaction energies closer to DMC, whereas CCSDT(Q) tends to get back closer to the CCSD(T) method, with the final value being closer to the CCSD(T) than the DMC value, reconfirming our previous results.

\subsection*{Acknowledgments}
Research at Weizmann was supported by the Minerva Foundation, Munich, Germany, and by an internal grant from the Uriel Arnon Memorial Fund for Artificial Intelligence in Materials Research.
Vladimir Fishman acknowledges a doctoral fellowship from WIS, as well as the Erasmus+ Exchange Programme of the European Union for a mobility grant to visit the University of Graz.
A. Daniel Boese was supported by the Weizmann Institute of Science as a Weston Visiting Scholar.
All calculations were carried out on the Faculty of Chemistry’s high-performance computing facility \texttt{CHEMFARM}, which is supported in part by the Ben May Center for Chemical Theory and Computation. The authors thank its administrators, Dr. Mark Vilensky and Andrei Vasilev, for their kind assistance.

\begin{suppinfo}
\begin{itemize}
    \item PDF document with full raw data, including additional discussions, extra methods used, data for further polyaromatic systems, can be found in the Supporting Information.
\item Cartesian coordinate (xyz) files of the structures used in this paper are given.
\end{itemize}
\end{suppinfo}

\bibliography{main}

\providecommand{\latin}[1]{#1}
\makeatletter
\providecommand{\doi}
  {\begingroup\let\do\@makeother\dospecials
  \catcode`\{=1 \catcode`\}=2 \doi@aux}
\providecommand{\doi@aux}[1]{\endgroup\texttt{#1}}
\makeatother
\providecommand*\mcitethebibliography{\thebibliography}
\csname @ifundefined\endcsname{endmcitethebibliography}  {\let\endmcitethebibliography\endthebibliography}{}
\begin{mcitethebibliography}{77}
\providecommand*\natexlab[1]{#1}
\providecommand*\mciteSetBstSublistMode[1]{}
\providecommand*\mciteSetBstMaxWidthForm[2]{}
\providecommand*\mciteBstWouldAddEndPuncttrue
  {\def\EndOfBibitem{\unskip.}}
\providecommand*\mciteBstWouldAddEndPunctfalse
  {\let\EndOfBibitem\relax}
\providecommand*\mciteSetBstMidEndSepPunct[3]{}
\providecommand*\mciteSetBstSublistLabelBeginEnd[3]{}
\providecommand*\EndOfBibitem{}
\mciteSetBstSublistMode{f}
\mciteSetBstMaxWidthForm{subitem}{(\alph{mcitesubitemcount})}
\mciteSetBstSublistLabelBeginEnd
  {\mcitemaxwidthsubitemform\space}
  {\relax}
  {\relax}

\bibitem[Saleh \latin{et~al.}(2012)Saleh, Gatti, Lo~Presti, and Contreras-Garc{\'\i}a]{saleh2012revealing}
Saleh,~G.; Gatti,~C.; Lo~Presti,~L.; Contreras-Garc{\'\i}a,~J. Revealing non-covalent interactions in molecular crystals through their experimental electron densities. \emph{Chem. Eur. J.} \textbf{2012}, \emph{18}, 15523--15536, DOI: \doi{10.1002/chem.201201290}\relax
\mciteBstWouldAddEndPuncttrue
\mciteSetBstMidEndSepPunct{\mcitedefaultmidpunct}
{\mcitedefaultendpunct}{\mcitedefaultseppunct}\relax
\EndOfBibitem
\bibitem[Hoja \latin{et~al.}(2017)Hoja, Reilly, and Tkatchenko]{hoja2017first}
Hoja,~J.; Reilly,~A.~M.; Tkatchenko,~A. First-principles modeling of molecular crystals: structures and stabilities, temperature and pressure. \emph{Wiley Interdiscip. Rev.: Comput. Mol. Sci.} \textbf{2017}, \emph{7}, e1294, DOI: \doi{10.1002/wcms.1294}\relax
\mciteBstWouldAddEndPuncttrue
\mciteSetBstMidEndSepPunct{\mcitedefaultmidpunct}
{\mcitedefaultendpunct}{\mcitedefaultseppunct}\relax
\EndOfBibitem
\bibitem[Storer and Hunter(2022)Storer, and Hunter]{storer2022surface}
Storer,~M.~C.; Hunter,~C.~A. The surface site interaction point approach to non-covalent interactions. \emph{Chem. Soc. Rev.} \textbf{2022}, \emph{51}, 10064--10082, DOI: \doi{10.1039/D2CS00701K}\relax
\mciteBstWouldAddEndPuncttrue
\mciteSetBstMidEndSepPunct{\mcitedefaultmidpunct}
{\mcitedefaultendpunct}{\mcitedefaultseppunct}\relax
\EndOfBibitem
\bibitem[Buaksuntear \latin{et~al.}(2022)Buaksuntear, Limarun, Suethao, and Smitthipong]{buaksuntear2022non}
Buaksuntear,~K.; Limarun,~P.; Suethao,~S.; Smitthipong,~W. Non-covalent interaction on the self-healing of mechanical properties in supramolecular polymers. \emph{International Journal of Molecular Sciences} \textbf{2022}, \emph{23}, 6902, DOI: \doi{10.3390/ijms23136902}\relax
\mciteBstWouldAddEndPuncttrue
\mciteSetBstMidEndSepPunct{\mcitedefaultmidpunct}
{\mcitedefaultendpunct}{\mcitedefaultseppunct}\relax
\EndOfBibitem
\bibitem[Chen \latin{et~al.}(2022)Chen, Peng, Peng, Zhang, and Zeng]{chen2022probing}
Chen,~J.; Peng,~Q.; Peng,~X.; Zhang,~H.; Zeng,~H. Probing and manipulating noncovalent interactions in functional polymeric systems. \emph{Chem. Rev.} \textbf{2022}, \emph{122}, 14594--14678, DOI: \doi{10.1021/acs.chemrev.2c00215}\relax
\mciteBstWouldAddEndPuncttrue
\mciteSetBstMidEndSepPunct{\mcitedefaultmidpunct}
{\mcitedefaultendpunct}{\mcitedefaultseppunct}\relax
\EndOfBibitem
\bibitem[Adhav and Saikrishnan(2023)Adhav, and Saikrishnan]{adhav2023realm}
Adhav,~V.~A.; Saikrishnan,~K. The realm of unconventional noncovalent interactions in proteins: their significance in structure and function. \emph{ACS Omega} \textbf{2023}, \emph{8}, 22268--22284, DOI: \doi{10.1021/acsomega.3c00205}\relax
\mciteBstWouldAddEndPuncttrue
\mciteSetBstMidEndSepPunct{\mcitedefaultmidpunct}
{\mcitedefaultendpunct}{\mcitedefaultseppunct}\relax
\EndOfBibitem
\bibitem[Tuncel(2011)]{tuncel2011non}
Tuncel,~D. Non-covalent interactions between carbon nanotubes and conjugated polymers. \emph{Nanoscale} \textbf{2011}, \emph{3}, 3545--3554, DOI: \doi{10.1039/C1NR10338E}\relax
\mciteBstWouldAddEndPuncttrue
\mciteSetBstMidEndSepPunct{\mcitedefaultmidpunct}
{\mcitedefaultendpunct}{\mcitedefaultseppunct}\relax
\EndOfBibitem
\bibitem[Rehman \latin{et~al.}(2015)Rehman, Sarwar, Husain, Ishqi, and Tabish]{rehman2015studying}
Rehman,~S.~U.; Sarwar,~T.; Husain,~M.~A.; Ishqi,~H.~M.; Tabish,~M. Studying non-covalent drug--DNA interactions. \emph{Archives of Biochemistry and Biophysics} \textbf{2015}, \emph{576}, 49--60, DOI: \doi{10.1016/j.abb.2015.04.004}\relax
\mciteBstWouldAddEndPuncttrue
\mciteSetBstMidEndSepPunct{\mcitedefaultmidpunct}
{\mcitedefaultendpunct}{\mcitedefaultseppunct}\relax
\EndOfBibitem
\bibitem[Raynal \latin{et~al.}(2014)Raynal, Ballester, Vidal-Ferran, and van Leeuwen]{raynal2014supramolecular}
Raynal,~M.; Ballester,~P.; Vidal-Ferran,~A.; van Leeuwen,~P.~W. Supramolecular catalysis. Part 1: non-covalent interactions as a tool for building and modifying homogeneous catalysts. \emph{Chem. Soc. Rev.} \textbf{2014}, \emph{43}, 1660--1733, DOI: \doi{10.1039/C3CS60027K}\relax
\mciteBstWouldAddEndPuncttrue
\mciteSetBstMidEndSepPunct{\mcitedefaultmidpunct}
{\mcitedefaultendpunct}{\mcitedefaultseppunct}\relax
\EndOfBibitem
\bibitem[Al-Hamdani \latin{et~al.}(2021)Al-Hamdani, Nagy, Zen, Barton, Kállay, Brandenburg, and Tkatchenko]{AlHamdani2021}
Al-Hamdani,~Y.~S.; Nagy,~P.~R.; Zen,~A.; Barton,~D.; Kállay,~M.; Brandenburg,~J.~G.; Tkatchenko,~A. Interactions between large molecules pose a puzzle for reference quantum mechanical methods. \emph{Nat. Commun.} \textbf{2021}, \emph{12}, 3927, DOI: \doi{10.1038/s41467-021-24119-3}\relax
\mciteBstWouldAddEndPuncttrue
\mciteSetBstMidEndSepPunct{\mcitedefaultmidpunct}
{\mcitedefaultendpunct}{\mcitedefaultseppunct}\relax
\EndOfBibitem
\bibitem[Reynolds \latin{et~al.}(1982)Reynolds, Ceperley, Alder, and Lester~Jr]{reynolds1982fixed}
Reynolds,~P.~J.; Ceperley,~D.~M.; Alder,~B.~J.; Lester~Jr,~W.~A. Fixed-node quantum Monte Carlo for molecules. \emph{J. Chem. Phys.} \textbf{1982}, \emph{77}, 5593--5603, DOI: \doi{10.1063/1.443766}\relax
\mciteBstWouldAddEndPuncttrue
\mciteSetBstMidEndSepPunct{\mcitedefaultmidpunct}
{\mcitedefaultendpunct}{\mcitedefaultseppunct}\relax
\EndOfBibitem
\bibitem[Della~Pia \latin{et~al.}(2025)Della~Pia, Shi, Al-Hamdani, Alf{\`e}, Anderson, Barborini, Benali, Casula, Drummond, Dubeck{\`y}, \latin{et~al.} others]{della2025fixed}
Della~Pia,~F.; Shi,~B.~X.; Al-Hamdani,~Y.~S.; Alf{\`e},~D.; Anderson,~T.~A.; Barborini,~M.; Benali,~A.; Casula,~M.; Drummond,~N.~D.; Dubeck{\`y},~M. \latin{et~al.}  Reproducibility of fixed-node diffusion Monte Carlo across diverse community codes: The case of water–methane dimer. \emph{J. Chem. Phys.} \textbf{2025}, \emph{163}, DOI: \doi{10.1063/5.0272974}\relax
\mciteBstWouldAddEndPuncttrue
\mciteSetBstMidEndSepPunct{\mcitedefaultmidpunct}
{\mcitedefaultendpunct}{\mcitedefaultseppunct}\relax
\EndOfBibitem
\bibitem[Raghavachari \latin{et~al.}(1989)Raghavachari, Trucks, Pople, and Head-Gordon]{Raghavachari1989}
Raghavachari,~K.; Trucks,~G.~W.; Pople,~J.~A.; Head-Gordon,~M. A Fifth-Order Perturbation Comparison of Electron Correlation Theories. \emph{Chem. Phys. Lett.} \textbf{1989}, \emph{157}, 479--483, DOI: \doi{10.1016/S0009-2614(89)87395-6}\relax
\mciteBstWouldAddEndPuncttrue
\mciteSetBstMidEndSepPunct{\mcitedefaultmidpunct}
{\mcitedefaultendpunct}{\mcitedefaultseppunct}\relax
\EndOfBibitem
\bibitem[Gyevi-Nagy \latin{et~al.}(2020)Gyevi-Nagy, K\'allay, and Nagy]{MPICCSDpT}
Gyevi-Nagy,~L.; K\'allay,~M.; Nagy,~P.~R. Integral-direct and parallel implementation of the {CCSD(T)} method: {A}lgorithmic developments and large-scale applications. \emph{J. Chem. Theory Comput.} \textbf{2020}, \emph{16}, 366, DOI: \doi{10.1021/acs.jctc.9b00957}\relax
\mciteBstWouldAddEndPuncttrue
\mciteSetBstMidEndSepPunct{\mcitedefaultmidpunct}
{\mcitedefaultendpunct}{\mcitedefaultseppunct}\relax
\EndOfBibitem
\bibitem[Pickard~IV \latin{et~al.}(2006)Pickard~IV, Griffith, Ferrara, Liptak, Kirschner, and Shields]{pickard2006ccsd}
Pickard~IV,~F.~C.; Griffith,~D.~R.; Ferrara,~S.~J.; Liptak,~M.~D.; Kirschner,~K.~N.; Shields,~G.~C. CCSD (T), W1, and other model chemistry predictions for gas-phase deprotonation reactions. \emph{Int. J. Quantum Chem.} \textbf{2006}, \emph{106}, 3122--3128, DOI: \doi{10.1002/qua.21105}\relax
\mciteBstWouldAddEndPuncttrue
\mciteSetBstMidEndSepPunct{\mcitedefaultmidpunct}
{\mcitedefaultendpunct}{\mcitedefaultseppunct}\relax
\EndOfBibitem
\bibitem[Fishman \latin{et~al.}(2025)Fishman, Lesiuk, Martin, and Boese]{fishman2025another}
Fishman,~V.; Lesiuk,~M.; Martin,~J.~M.; Boese,~A.~D. Another angle on benchmarking noncovalent interactions. \emph{J. Chem. Theory Comput.} \textbf{2025}, \emph{21}, 2311--2324, DOI: \doi{10.1021/acs.jctc.4c01512}\relax
\mciteBstWouldAddEndPuncttrue
\mciteSetBstMidEndSepPunct{\mcitedefaultmidpunct}
{\mcitedefaultendpunct}{\mcitedefaultseppunct}\relax
\EndOfBibitem
\bibitem[Lesiuk(2022)]{Lesiuk22}
Lesiuk,~M. When Gold Is Not Enough: Platinum Standard of Quantum Chemistry with $N^7$ Cost. \emph{J. Chem. Theory Comput.} \textbf{2022}, \emph{18}, 6537--6556, DOI: \doi{10.1021/acs.jctc.2c00460}\relax
\mciteBstWouldAddEndPuncttrue
\mciteSetBstMidEndSepPunct{\mcitedefaultmidpunct}
{\mcitedefaultendpunct}{\mcitedefaultseppunct}\relax
\EndOfBibitem
\bibitem[Bomble \latin{et~al.}(2005)Bomble, Stanton, K{\'a}llay, and Gauss]{bomble2005coupled}
Bomble,~Y.~J.; Stanton,~J.~F.; K{\'a}llay,~M.; Gauss,~J. Coupled-cluster methods including noniterative corrections for quadruple excitations. \emph{J. Chem. Phys.} \textbf{2005}, \emph{123}, 054101, DOI: \doi{10.1063/1.1950567}\relax
\mciteBstWouldAddEndPuncttrue
\mciteSetBstMidEndSepPunct{\mcitedefaultmidpunct}
{\mcitedefaultendpunct}{\mcitedefaultseppunct}\relax
\EndOfBibitem
\bibitem[Boese(2013)]{boese2013assesment}
Boese,~A.~D. Assessment of Coupled Cluster Theory and more Approximate Methods for Hydrogen Bonded Systems. \emph{J. Chem. Theory Comput.} \textbf{2013}, \emph{9}, 4403--4413, DOI: \doi{10.1021/ct400558w}\relax
\mciteBstWouldAddEndPuncttrue
\mciteSetBstMidEndSepPunct{\mcitedefaultmidpunct}
{\mcitedefaultendpunct}{\mcitedefaultseppunct}\relax
\EndOfBibitem
\bibitem[Lambie \latin{et~al.}(2025)Lambie, Kats, Usvyat, and Alavi]{Lambie2025CCSDTDispersion}
Lambie,~S.; Kats,~D.; Usvyat,~D.; Alavi,~A. On the applicability of CCSD(T) for dispersion interactions in large conjugated systems. \emph{J. Chem. Phys.} \textbf{2025}, \emph{162}, 114112, DOI: \doi{10.1063/5.0246763}\relax
\mciteBstWouldAddEndPuncttrue
\mciteSetBstMidEndSepPunct{\mcitedefaultmidpunct}
{\mcitedefaultendpunct}{\mcitedefaultseppunct}\relax
\EndOfBibitem
\bibitem[Briccolani-Bandini \latin{et~al.}(2026)Briccolani-Bandini, Gnnanapareddy, Labat, Br{\'e}mond, Sancho-Garc{\'\i}a, Otero-de-la Roza, Scalmani, Frisch, Cardini, DiLabio, and Adamo]{briccolani2026weak}
Briccolani-Bandini,~L.; Gnnanapareddy,~B.; Labat,~F.; Br{\'e}mond,~E.; Sancho-Garc{\'\i}a,~J.~C.; Otero-de-la Roza,~A.; Scalmani,~G.; Frisch,~M.~J.; Cardini,~G.; DiLabio,~G.~A. \latin{et~al.}  Weak Noncovalent Interactions in Nonequilibrium Structures: How Good Are the Dispersion Corrections? \emph{J. Phys. Chem. Lett.} \textbf{2026}, \emph{17}, 703--711, DOI: \doi{10.1021/acs.jpclett.5c02634}\relax
\mciteBstWouldAddEndPuncttrue
\mciteSetBstMidEndSepPunct{\mcitedefaultmidpunct}
{\mcitedefaultendpunct}{\mcitedefaultseppunct}\relax
\EndOfBibitem
\bibitem[Semidalas \latin{et~al.}(2025)Semidalas, Boese, and Martin]{Semidalas2025}
Semidalas,~E.; Boese,~A.~D.; Martin,~J.~M. Post-CCSD(T) corrections in the S66 noncovalent interactions benchmark. \textbf{2025}, \emph{863}, 141874, DOI: \doi{10.1016/j.cplett.2025.141874}\relax
\mciteBstWouldAddEndPuncttrue
\mciteSetBstMidEndSepPunct{\mcitedefaultmidpunct}
{\mcitedefaultendpunct}{\mcitedefaultseppunct}\relax
\EndOfBibitem
\bibitem[Herbert(2021)]{herbert2021neat}
Herbert,~J.~M. Neat, Simple, and Wrong: Debunking Electrostatic Fallacies Regarding Noncovalent Interactions. \emph{J. Phys. Chem. A} \textbf{2021}, \emph{125}, 7125--7137, DOI: \doi{10.1021/acs.jpca.1c05962}\relax
\mciteBstWouldAddEndPuncttrue
\mciteSetBstMidEndSepPunct{\mcitedefaultmidpunct}
{\mcitedefaultendpunct}{\mcitedefaultseppunct}\relax
\EndOfBibitem
\bibitem[Hobza \latin{et~al.}(1996)Hobza, Selzle, and Schlag]{Hobza1996BenzeneDimerIsoenergetic}
Hobza,~P.; Selzle,~H.~L.; Schlag,~E.~W. Potential Energy Surface for the Benzene Dimer. Results of ab Initio {CCSD(T)} Calculations Show Two Nearly Isoenergetic Structures: {T}-Shaped and Parallel-Displaced. \emph{J. Phys. Chem.} \textbf{1996}, \emph{100}, 18790--18794, DOI: \doi{10.1021/jp961239y}\relax
\mciteBstWouldAddEndPuncttrue
\mciteSetBstMidEndSepPunct{\mcitedefaultmidpunct}
{\mcitedefaultendpunct}{\mcitedefaultseppunct}\relax
\EndOfBibitem
\bibitem[Law \latin{et~al.}(1984)Law, Schauer, and Bernstein]{law1984dimers}
Law,~K.; Schauer,~M.; Bernstein,~E.~R. Dimers of aromatic molecules: (Benzene)2, (toluene)2, and benzene–toluene. \emph{J. Chem. Phys.} \textbf{1984}, \emph{81}, 4871--4882, DOI: \doi{10.1063/1.447514}\relax
\mciteBstWouldAddEndPuncttrue
\mciteSetBstMidEndSepPunct{\mcitedefaultmidpunct}
{\mcitedefaultendpunct}{\mcitedefaultseppunct}\relax
\EndOfBibitem
\bibitem[Matsokin \latin{et~al.}(2026)Matsokin, Kalinin, Buchwald, Werner, Fink, Fink, and Sharapa]{matsokin2026revisiting}
Matsokin,~N.~A.; Kalinin,~M.; Buchwald,~A.; Werner,~H.-J.; Fink,~R.~F.; Fink,~K.; Sharapa,~D.~I. Revisiting acene dimers: A comprehensive theoretical study of a less explored conformer. \emph{J. Chem. Phys.} \textbf{2026}, \emph{164}, DOI: \doi{10.1063/5.0324531}\relax
\mciteBstWouldAddEndPuncttrue
\mciteSetBstMidEndSepPunct{\mcitedefaultmidpunct}
{\mcitedefaultendpunct}{\mcitedefaultseppunct}\relax
\EndOfBibitem
\bibitem[Hohenberg and Kohn(1964)Hohenberg, and Kohn]{Hohenberg1964}
Hohenberg,~P.; Kohn,~W. Inhomogeneous Electron Gas. \emph{Phys. Rev.} \textbf{1964}, \emph{136}, B864--B871, DOI: \doi{10.1103/PhysRev.136.B864}\relax
\mciteBstWouldAddEndPuncttrue
\mciteSetBstMidEndSepPunct{\mcitedefaultmidpunct}
{\mcitedefaultendpunct}{\mcitedefaultseppunct}\relax
\EndOfBibitem
\bibitem[Kohn and Sham(1965)Kohn, and Sham]{Kohn1965}
Kohn,~W.; Sham,~L.~J. Self-Consistent Equations Including Exchange and Correlation Effects. \emph{Phys. Rev.} \textbf{1965}, \emph{140}, A1133--A1138, DOI: \doi{10.1103/PhysRev.140.A1133}\relax
\mciteBstWouldAddEndPuncttrue
\mciteSetBstMidEndSepPunct{\mcitedefaultmidpunct}
{\mcitedefaultendpunct}{\mcitedefaultseppunct}\relax
\EndOfBibitem
\bibitem[Mardirossian and Head-Gordon(2015)Mardirossian, and Head-Gordon]{Mardirossian2015}
Mardirossian,~N.; Head-Gordon,~M. Mapping the Genome of Meta-Generalized Gradient Approximation Density Functionals: The Search for B97M-V. \emph{J. Chem. Phys.} \textbf{2015}, \emph{142}, 074111, DOI: \doi{10.1063/1.4907719}\relax
\mciteBstWouldAddEndPuncttrue
\mciteSetBstMidEndSepPunct{\mcitedefaultmidpunct}
{\mcitedefaultendpunct}{\mcitedefaultseppunct}\relax
\EndOfBibitem
\bibitem[Weigend and Ahlrichs(2005)Weigend, and Ahlrichs]{Weigend2005}
Weigend,~F.; Ahlrichs,~R. Balanced basis sets of split valence, triple zeta valence and quadruple zeta valence quality for H to Rn: Design and assessment of accuracy. \emph{Phys. Chem. Chem. Phys.} \textbf{2005}, \emph{7}, 3297--3305, DOI: \doi{10.1039/B508541A}\relax
\mciteBstWouldAddEndPuncttrue
\mciteSetBstMidEndSepPunct{\mcitedefaultmidpunct}
{\mcitedefaultendpunct}{\mcitedefaultseppunct}\relax
\EndOfBibitem
\bibitem[Neese \latin{et~al.}(2020)Neese, Wennmohs, Becker, and Riplinger]{neese2020orca}
Neese,~F.; Wennmohs,~F.; Becker,~U.; Riplinger,~C. The ORCA quantum chemistry program package. \emph{J. Chem. Phys.} \textbf{2020}, \emph{152}, 224108, DOI: \doi{10.1063/5.0004608}\relax
\mciteBstWouldAddEndPuncttrue
\mciteSetBstMidEndSepPunct{\mcitedefaultmidpunct}
{\mcitedefaultendpunct}{\mcitedefaultseppunct}\relax
\EndOfBibitem
\bibitem[Epifanovsky \latin{et~al.}(2013)Epifanovsky, Zuev, Feng, Khistyaev, Shao, and Krylov]{epifanovsky2013general}
Epifanovsky,~E.; Zuev,~D.; Feng,~X.; Khistyaev,~K.; Shao,~Y.; Krylov,~A.~I. General implementation of the resolution-of-the-identity and Cholesky representations of electron repulsion integrals within coupled-cluster and equation-of-motion methods: Theory and benchmarks. \emph{J. Chem. Phys.} \textbf{2013}, \emph{139}, 134105, DOI: \doi{10.1063/1.4820484}\relax
\mciteBstWouldAddEndPuncttrue
\mciteSetBstMidEndSepPunct{\mcitedefaultmidpunct}
{\mcitedefaultendpunct}{\mcitedefaultseppunct}\relax
\EndOfBibitem
\bibitem[Shen \latin{et~al.}(2019)Shen, Zhu, Zhang, and Scheffler]{shen2019massive}
Shen,~T.; Zhu,~Z.; Zhang,~I.~Y.; Scheffler,~M. Massive-parallel implementation of the resolution-of-identity coupled-cluster approaches in the numeric atom-centered orbital framework for molecular systems. \emph{J. Chem. Theory Comput.} \textbf{2019}, \emph{15}, 4721--4734, DOI: \doi{10.1021/acs.jctc.8b01294}\relax
\mciteBstWouldAddEndPuncttrue
\mciteSetBstMidEndSepPunct{\mcitedefaultmidpunct}
{\mcitedefaultendpunct}{\mcitedefaultseppunct}\relax
\EndOfBibitem
\bibitem[{K\'allay} \latin{et~al.}(2020){K\'allay}, Nagy, Mester, Rolik, Samu, Csontos, {Cs\'oka}, {Szab\'o}, Gyevi-Nagy, {H\'egely}, {Ladj\'anszki}, Szegedy, {Lad\'oczki}, Petrov, Farkas, Mezei, and Ganyecz]{MRCC}
{K\'allay},~M.; Nagy,~P.~R.; Mester,~D.; Rolik,~Z.; Samu,~G.; Csontos,~J.; {Cs\'oka},~J.; {Szab\'o},~P.~B.; Gyevi-Nagy,~L.; {H\'egely},~B. \latin{et~al.}  The {MRCC} program system: {A}ccurate quantum chemistry from water to proteins. \emph{J. Chem. Phys.} \textbf{2020}, \emph{152}, 074107, DOI: \doi{10.1063/1.5142048}\relax
\mciteBstWouldAddEndPuncttrue
\mciteSetBstMidEndSepPunct{\mcitedefaultmidpunct}
{\mcitedefaultendpunct}{\mcitedefaultseppunct}\relax
\EndOfBibitem
\bibitem[Nagy and Kállay(2019)Nagy, and Kállay]{Nagy2019}
Nagy,~P.~R.; Kállay,~M. Approaching the Basis Set Limit of CCSD(T) Energies for Large Molecules with Local Natural Orbital Coupled-Cluster Methods. \emph{J. Chem. Theory Comput.} \textbf{2019}, \emph{15}, 5275--5298, DOI: \doi{10.1021/acs.jctc.9b00538}\relax
\mciteBstWouldAddEndPuncttrue
\mciteSetBstMidEndSepPunct{\mcitedefaultmidpunct}
{\mcitedefaultendpunct}{\mcitedefaultseppunct}\relax
\EndOfBibitem
\bibitem[Nagy(2024)]{local_corr_review2024}
Nagy,~P.~R. State-of-the-art local correlation methods enable accurate and affordable gold standard quantum chemistry up to a few hundred atoms. \emph{Chem. Sci.} \textbf{2024}, \emph{15}, 14556, DOI: \doi{10.1039/D4SC04755A}\relax
\mciteBstWouldAddEndPuncttrue
\mciteSetBstMidEndSepPunct{\mcitedefaultmidpunct}
{\mcitedefaultendpunct}{\mcitedefaultseppunct}\relax
\EndOfBibitem
\bibitem[Riplinger and Neese(2013)Riplinger, and Neese]{riplinger2013efficient}
Riplinger,~C.; Neese,~F. An efficient and near linear scaling pair natural orbital based local coupled cluster method. \emph{J. Chem. Phys.} \textbf{2013}, \emph{138}, 034106, DOI: \doi{10.1063/1.4773581}\relax
\mciteBstWouldAddEndPuncttrue
\mciteSetBstMidEndSepPunct{\mcitedefaultmidpunct}
{\mcitedefaultendpunct}{\mcitedefaultseppunct}\relax
\EndOfBibitem
\bibitem[Riplinger \latin{et~al.}(2013)Riplinger, Sandhoefer, Hansen, and Neese]{riplinger2013natural}
Riplinger,~C.; Sandhoefer,~B.; Hansen,~A.; Neese,~F. Natural triple excitations in local coupled cluster calculations with pair natural orbitals. \emph{J. Chem. Phys.} \textbf{2013}, \emph{139}, 134101, DOI: \doi{10.1063/1.4821834}\relax
\mciteBstWouldAddEndPuncttrue
\mciteSetBstMidEndSepPunct{\mcitedefaultmidpunct}
{\mcitedefaultendpunct}{\mcitedefaultseppunct}\relax
\EndOfBibitem
\bibitem[Riplinger \latin{et~al.}(2016)Riplinger, Pinski, Becker, Valeev, and Neese]{riplinger2016sparse}
Riplinger,~C.; Pinski,~P.; Becker,~U.; Valeev,~E.~F.; Neese,~F. Sparse maps—A systematic infrastructure for reduced-scaling electronic structure methods. II. Linear scaling domain based pair natural orbital coupled cluster theory. \emph{J. Chem. Phys.} \textbf{2016}, \emph{144}, DOI: \doi{10.1063/1.4939030}\relax
\mciteBstWouldAddEndPuncttrue
\mciteSetBstMidEndSepPunct{\mcitedefaultmidpunct}
{\mcitedefaultendpunct}{\mcitedefaultseppunct}\relax
\EndOfBibitem
\bibitem[Guo \latin{et~al.}(2018)Guo, Riplinger, Becker, Liakos, Minenkov, Cavallo, and Neese]{guo2018communication}
Guo,~Y.; Riplinger,~C.; Becker,~U.; Liakos,~D.~G.; Minenkov,~Y.; Cavallo,~L.; Neese,~F. Communication: An improved linear scaling perturbative triples correction for the domain based local pair-natural orbital based singles and doubles coupled cluster method [DLPNO-CCSD (T)]. \emph{J. Chem. Phys.} \textbf{2018}, \emph{148}, DOI: \doi{10.1063/1.5011798}\relax
\mciteBstWouldAddEndPuncttrue
\mciteSetBstMidEndSepPunct{\mcitedefaultmidpunct}
{\mcitedefaultendpunct}{\mcitedefaultseppunct}\relax
\EndOfBibitem
\bibitem[Ma and Werner(2018)Ma, and Werner]{ma2018explicitly}
Ma,~Q.; Werner,~H.-J. Explicitly correlated local coupled-cluster methods using pair natural orbitals. \emph{Wiley Interdisciplinary Reviews: Computational Molecular Science} \textbf{2018}, \emph{8}, e1371, DOI: \doi{10.1002/wcms.1371}\relax
\mciteBstWouldAddEndPuncttrue
\mciteSetBstMidEndSepPunct{\mcitedefaultmidpunct}
{\mcitedefaultendpunct}{\mcitedefaultseppunct}\relax
\EndOfBibitem
\bibitem[Werner \latin{et~al.}(2020)Werner, Knowles, Knizia, Manby, and {Sch{\"u}tz}]{Werner2020}
Werner,~H.-J.; Knowles,~P.~J.; Knizia,~G.; Manby,~F.~R.; {Sch{\"u}tz},~M. MOLPRO: a general-purpose quantum chemistry program package. \emph{WIREs Computational Molecular Science} \textbf{2020}, \emph{10}, e1427, DOI: \doi{10.1002/wcms.1427}\relax
\mciteBstWouldAddEndPuncttrue
\mciteSetBstMidEndSepPunct{\mcitedefaultmidpunct}
{\mcitedefaultendpunct}{\mcitedefaultseppunct}\relax
\EndOfBibitem
\bibitem[Stanton and Gauss(1996)Stanton, and Gauss]{stanton1996simple}
Stanton,~J.~F.; Gauss,~J. A simple correction to final state energies of doublet radicals described by equation-of-motion coupled cluster theory in the singles and doubles approximation. \emph{Theor. Chem. Acc.} \textbf{1996}, \emph{93}, 303--313, DOI: \doi{10.1007/s002140050154}\relax
\mciteBstWouldAddEndPuncttrue
\mciteSetBstMidEndSepPunct{\mcitedefaultmidpunct}
{\mcitedefaultendpunct}{\mcitedefaultseppunct}\relax
\EndOfBibitem
\bibitem[Crawford and Stanton(1998)Crawford, and Stanton]{crawford1998investigation}
Crawford,~T.~D.; Stanton,~J.~F. Investigation of an asymmetric triple-excitation correction for coupled-cluster energies. \emph{Int. J. Quantum Chem.} \textbf{1998}, \emph{70}, 601--611, DOI: \doi{10.1002/(sici)1097-461x(1998)70:4/5<601::aid-qua6>3.0.co;2-z}\relax
\mciteBstWouldAddEndPuncttrue
\mciteSetBstMidEndSepPunct{\mcitedefaultmidpunct}
{\mcitedefaultendpunct}{\mcitedefaultseppunct}\relax
\EndOfBibitem
\bibitem[Kucharski and Bartlett(1998)Kucharski, and Bartlett]{kucharski1998noniterative}
Kucharski,~S.~A.; Bartlett,~R.~J. Noniterative energy corrections through fifth-order to the coupled cluster singles and doubles method. \emph{J. Chem. Phys.} \textbf{1998}, \emph{108}, 5243--5254, DOI: \doi{10.1063/1.475961}\relax
\mciteBstWouldAddEndPuncttrue
\mciteSetBstMidEndSepPunct{\mcitedefaultmidpunct}
{\mcitedefaultendpunct}{\mcitedefaultseppunct}\relax
\EndOfBibitem
\bibitem[Kucharski and Bartlett(1998)Kucharski, and Bartlett]{kucharski1998sixth}
Kucharski,~S.~A.; Bartlett,~R.~J. Sixth-order energy corrections with converged coupled cluster singles and doubles amplitudes. \emph{J. Chem. Phys.} \textbf{1998}, \emph{108}, 5255--5264, DOI: \doi{10.1063/1.475962}\relax
\mciteBstWouldAddEndPuncttrue
\mciteSetBstMidEndSepPunct{\mcitedefaultmidpunct}
{\mcitedefaultendpunct}{\mcitedefaultseppunct}\relax
\EndOfBibitem
\bibitem[Matthews \latin{et~al.}(2020)Matthews, Cheng, Harding, Lipparini, Stopkowicz, Jagau, Szalay, Gauss, and Stanton]{CFOUR}
Matthews,~D.~A.; Cheng,~L.; Harding,~M.~E.; Lipparini,~F.; Stopkowicz,~S.; Jagau,~T.-C.; Szalay,~P.~G.; Gauss,~J.; Stanton,~J.~F. {Coupled-cluster techniques for computational chemistry: The CFOUR program package}. \emph{J. Chem. Phys.} \textbf{2020}, \emph{152}, 214108, DOI: \doi{10.1063/5.0004837}\relax
\mciteBstWouldAddEndPuncttrue
\mciteSetBstMidEndSepPunct{\mcitedefaultmidpunct}
{\mcitedefaultendpunct}{\mcitedefaultseppunct}\relax
\EndOfBibitem
\bibitem[Stanton(1997)]{stanton1997ccsd}
Stanton,~J.~F. Why {CCSD(T)} works: a different perspective. \emph{Chem. Phys. Lett.} \textbf{1997}, \emph{281}, 130--134, DOI: \doi{10.1016/S0009-2614(97)01144-5}\relax
\mciteBstWouldAddEndPuncttrue
\mciteSetBstMidEndSepPunct{\mcitedefaultmidpunct}
{\mcitedefaultendpunct}{\mcitedefaultseppunct}\relax
\EndOfBibitem
\bibitem[Noga \latin{et~al.}(1987)Noga, Bartlett, and Urban]{noga1987towards}
Noga,~J.; Bartlett,~R.~J.; Urban,~M. Towards a full CCSDT model for electron correlation. CCSDT-n models. \emph{Chem. Phys. Lett.} \textbf{1987}, \emph{134}, 126--132, DOI: \doi{10.1016/0009-2614(87)87107-5}\relax
\mciteBstWouldAddEndPuncttrue
\mciteSetBstMidEndSepPunct{\mcitedefaultmidpunct}
{\mcitedefaultendpunct}{\mcitedefaultseppunct}\relax
\EndOfBibitem
\bibitem[He \latin{et~al.}(2001)He, He, and Cremer]{He2001comparisonCCSDTs}
He,~Y.; He,~Z.; Cremer,~D. Comparison of CCSDT-n methods with coupled-cluster theory with single and double excitations and coupled-cluster theory with single, double, and triple excitations in terms of many-body perturbation theory--what is the most effective triple-excitation method? \emph{Theor. Chem. Acc.} \textbf{2001}, \emph{105}, 182--196, DOI: \doi{10.1007/s002140000196}\relax
\mciteBstWouldAddEndPuncttrue
\mciteSetBstMidEndSepPunct{\mcitedefaultmidpunct}
{\mcitedefaultendpunct}{\mcitedefaultseppunct}\relax
\EndOfBibitem
\bibitem[M{\o}ller and Plesset(1934)M{\o}ller, and Plesset]{moller1934note}
M{\o}ller,~C.; Plesset,~M.~S. Note on an approximation treatment for many-electron systems. \emph{Phys. Rev.} \textbf{1934}, \emph{46}, 618, DOI: \doi{10.1103/PhysRev.46.618}\relax
\mciteBstWouldAddEndPuncttrue
\mciteSetBstMidEndSepPunct{\mcitedefaultmidpunct}
{\mcitedefaultendpunct}{\mcitedefaultseppunct}\relax
\EndOfBibitem
\bibitem[Loipersberger \latin{et~al.}(2021)Loipersberger, Bertels, Lee, and Head-Gordon]{Loipersberger2021}
Loipersberger,~M.; Bertels,~L.~W.; Lee,~J.; Head-Gordon,~M. Exploring the Limits of Second- and Third-Order Møller–Plesset Perturbation Theories for Noncovalent Interactions: Revisiting MP2.5 and Assessing the Importance of Regularization and Reference Orbitals. \emph{J. Chem. Theory Comput.} \textbf{2021}, \emph{17}, 5582–5599, DOI: \doi{10.1021/acs.jctc.1c00469}\relax
\mciteBstWouldAddEndPuncttrue
\mciteSetBstMidEndSepPunct{\mcitedefaultmidpunct}
{\mcitedefaultendpunct}{\mcitedefaultseppunct}\relax
\EndOfBibitem
\bibitem[Shao \latin{et~al.}(2015)Shao, Gan, Epifanovsky, Gilbert, Wormit, Kussmann, Lange, He, Diedenhofen, Walker, Vreven, Jr., Zuev, Jr., Adamson, Austin, Dutoi, Simmonett, Villalobos, Zhang, Giorgioni, Williams-Young, Ding, {de Oliveira}, Tao, {Ulerich}, {Krylov}, {Valeev}, {Schaefer}, {Schaefer}, {Paul}, and {Head-Gordon}]{Shao2015}
Shao,~Y.; Gan,~Z.; Epifanovsky,~E.; Gilbert,~A.~M.; Wormit,~M.; Kussmann,~J.; Lange,~A.~W.; He,~X.; Diedenhofen,~T. G.~W.; Walker,~M.~A. \latin{et~al.}  Advances in Molecular Quantum Chemistry Contained in the Q-Chem 4 Program Package. \emph{Mol. Phys.} \textbf{2015}, \emph{113}, 184--215, DOI: \doi{10.1080/00268976.2014.952696}\relax
\mciteBstWouldAddEndPuncttrue
\mciteSetBstMidEndSepPunct{\mcitedefaultmidpunct}
{\mcitedefaultendpunct}{\mcitedefaultseppunct}\relax
\EndOfBibitem
\bibitem[Dunning~Jr(1989)]{dunning1989gaussian}
Dunning~Jr,~T.~H. Gaussian basis sets for use in correlated molecular calculations. I. The atoms boron through neon and hydrogen. \emph{J. Chem. Phys.} \textbf{1989}, \emph{90}, 1007--1023, DOI: \doi{10.1063/1.456153}\relax
\mciteBstWouldAddEndPuncttrue
\mciteSetBstMidEndSepPunct{\mcitedefaultmidpunct}
{\mcitedefaultendpunct}{\mcitedefaultseppunct}\relax
\EndOfBibitem
\bibitem[Woon and Dunning~Jr(1993)Woon, and Dunning~Jr]{woon1993gaussian}
Woon,~D.~E.; Dunning~Jr,~T.~H. Gaussian basis sets for use in correlated molecular calculations. III. The atoms aluminum through argon. \emph{J. Chem. Phys.} \textbf{1993}, \emph{98}, 1358--1371, DOI: \doi{10.1063/1.464303}\relax
\mciteBstWouldAddEndPuncttrue
\mciteSetBstMidEndSepPunct{\mcitedefaultmidpunct}
{\mcitedefaultendpunct}{\mcitedefaultseppunct}\relax
\EndOfBibitem
\bibitem[Martin \latin{et~al.}(2022)Martin, Santra, and Semidalas]{martin2022exchange}
Martin,~J. M.~L.; Santra,~G.; Semidalas,~E. An exchange-based diagnostic for static correlation. \emph{AIP Conf. Proc.} \textbf{2022}, \emph{2611}, 020014, DOI: \doi{10.1063/5.0119280}\relax
\mciteBstWouldAddEndPuncttrue
\mciteSetBstMidEndSepPunct{\mcitedefaultmidpunct}
{\mcitedefaultendpunct}{\mcitedefaultseppunct}\relax
\EndOfBibitem
\bibitem[Tao \latin{et~al.}(2003)Tao, Perdew, Staroverov, and Scuseria]{TPSS-functional}
Tao,~J.; Perdew,~J.~P.; Staroverov,~V.~N.; Scuseria,~G.~E. Climbing the Density Functional Ladder: Nonempirical Meta--Generalized Gradient Approximation Designed for Molecules and Solids. \emph{Phys. Rev. Lett.} \textbf{2003}, \emph{91}, 146401, DOI: \doi{10.1103/PhysRevLett.91.146401}\relax
\mciteBstWouldAddEndPuncttrue
\mciteSetBstMidEndSepPunct{\mcitedefaultmidpunct}
{\mcitedefaultendpunct}{\mcitedefaultseppunct}\relax
\EndOfBibitem
\bibitem[Lee and Taylor(1989)Lee, and Taylor]{lee1989diagnostic}
Lee,~T.~J.; Taylor,~P.~R. A diagnostic for determining the quality of single-reference electron correlation methods. \emph{Int. J. Quantum Chem.} \textbf{1989}, \emph{36}, 199--207, DOI: \doi{https://doi.org/10.1002/qua.560360824}\relax
\mciteBstWouldAddEndPuncttrue
\mciteSetBstMidEndSepPunct{\mcitedefaultmidpunct}
{\mcitedefaultendpunct}{\mcitedefaultseppunct}\relax
\EndOfBibitem
\bibitem[Lee \latin{et~al.}(1989)Lee, Rice, Scuseria, and Schaefer]{lee1989theoretical}
Lee,~T.~J.; Rice,~J.~E.; Scuseria,~G.~E.; Schaefer,~H.~F. Theoretical investigations of molecules composed only of fluorine, oxygen and nitrogen: determination of the equilibrium structures of FOOF,(NO) 2 and FNNF and the transition state structure for FNNF cis-trans isomerization. \emph{Theor. Chem. Acc.} \textbf{1989}, \emph{75}, 81--98, DOI: \doi{10.1007/BF00527711}\relax
\mciteBstWouldAddEndPuncttrue
\mciteSetBstMidEndSepPunct{\mcitedefaultmidpunct}
{\mcitedefaultendpunct}{\mcitedefaultseppunct}\relax
\EndOfBibitem
\bibitem[Langhoff and Bauschlicher~Jr.(1988)Langhoff, and Bauschlicher~Jr.]{langhoff1988ab}
Langhoff,~S.~R.; Bauschlicher~Jr.,~C.~W. Ab initio studies of transition metal systems. \emph{Annu. Rev. Phys. Chem.} \textbf{1988}, \emph{39}, 181--212, DOI: \doi{10.1146/annurev.pc.39.100188.001145}\relax
\mciteBstWouldAddEndPuncttrue
\mciteSetBstMidEndSepPunct{\mcitedefaultmidpunct}
{\mcitedefaultendpunct}{\mcitedefaultseppunct}\relax
\EndOfBibitem
\bibitem[\v{R}ez\'{a}\v{c} \latin{et~al.}(2011)\v{R}ez\'{a}\v{c}, Riley, and Hobza]{rezac2011s66}
\v{R}ez\'{a}\v{c},~J.; Riley,~K.~E.; Hobza,~P. S66: A Well-balanced Database of Benchmark Interaction Energies Relevant to Biomolecular Structures. \emph{J. Chem. Theory Comput.} \textbf{2011}, \emph{7}, 2427--2438, DOI: \doi{10.1021/ct2002946}\relax
\mciteBstWouldAddEndPuncttrue
\mciteSetBstMidEndSepPunct{\mcitedefaultmidpunct}
{\mcitedefaultendpunct}{\mcitedefaultseppunct}\relax
\EndOfBibitem
\bibitem[Masumian and Boese(2023)Masumian, and Boese]{Masumian2023}
Masumian,~E.; Boese,~A.~D. Benchmarking Swaths of Intermolecular Interaction Components with Symmetry-Adapted Perturbation Theory. \emph{J. Chem. Theory Comput.} \textbf{2023}, \emph{20}, 30–48, DOI: \doi{10.1021/acs.jctc.3c00801}\relax
\mciteBstWouldAddEndPuncttrue
\mciteSetBstMidEndSepPunct{\mcitedefaultmidpunct}
{\mcitedefaultendpunct}{\mcitedefaultseppunct}\relax
\EndOfBibitem
\bibitem[Saitow \latin{et~al.}(2017)Saitow, Becker, Riplinger, Valeev, and Neese]{saitow2017new}
Saitow,~M.; Becker,~U.; Riplinger,~C.; Valeev,~E.~F.; Neese,~F. A new near-linear scaling, efficient and accurate, open-shell domain-based local pair natural orbital coupled cluster singles and doubles theory. \emph{J. Chem. Phys.} \textbf{2017}, \emph{146}\relax
\mciteBstWouldAddEndPuncttrue
\mciteSetBstMidEndSepPunct{\mcitedefaultmidpunct}
{\mcitedefaultendpunct}{\mcitedefaultseppunct}\relax
\EndOfBibitem
\bibitem[Gray and Herbert(2024)Gray, and Herbert]{gray2024assessing}
Gray,~M.; Herbert,~J.~M. Assessing the domain-based local pair natural orbital (DLPNO) approximation for non-covalent interactions in sizable supramolecular complexes. \emph{J. Chem. Phys.} \textbf{2024}, \emph{161}\relax
\mciteBstWouldAddEndPuncttrue
\mciteSetBstMidEndSepPunct{\mcitedefaultmidpunct}
{\mcitedefaultendpunct}{\mcitedefaultseppunct}\relax
\EndOfBibitem
\bibitem[Semidalas and Martin(2022)Semidalas, and Martin]{semidalas2022mobh35}
Semidalas,~E.; Martin,~J. M.~L. {The {MOBH35} metal--organic barrier heights reconsidered: Performance of local-orbital coupled cluster approaches in different static correlation regimes}. \emph{J. Chem. Theory Comput.} \textbf{2022}, \emph{18}, 883--898, DOI: \doi{10.1021/acs.jctc.1c01126}\relax
\mciteBstWouldAddEndPuncttrue
\mciteSetBstMidEndSepPunct{\mcitedefaultmidpunct}
{\mcitedefaultendpunct}{\mcitedefaultseppunct}\relax
\EndOfBibitem
\bibitem[Semidalas \latin{et~al.}(2025)Semidalas, Boese, and Martin]{jmlm336}
Semidalas,~E.; Boese,~A.~D.; Martin,~J.~M. Post-CCSD (T) corrections in the S66 noncovalent interactions benchmark. \emph{Chem. Phys. Lett.} \textbf{2025}, \emph{863}, 141874, DOI: \doi{10.1016/j.cplett.2025.141874}\relax
\mciteBstWouldAddEndPuncttrue
\mciteSetBstMidEndSepPunct{\mcitedefaultmidpunct}
{\mcitedefaultendpunct}{\mcitedefaultseppunct}\relax
\EndOfBibitem
\bibitem[Boese(2015)]{boese20152}
Boese,~A.~D. Density Functional Theory and Hydrogen Bonds: Are We There Yet? \emph{ChemPhysChem} \textbf{2015}, \emph{16}, 978--985, DOI: \doi{https://doi.org/10.1002/cphc.201402786}\relax
\mciteBstWouldAddEndPuncttrue
\mciteSetBstMidEndSepPunct{\mcitedefaultmidpunct}
{\mcitedefaultendpunct}{\mcitedefaultseppunct}\relax
\EndOfBibitem
\bibitem[Sch{\"a}fer \latin{et~al.}(2025)Sch{\"a}fer, Irmler, Gallo, and Gr{\"u}neis]{schafer2025understanding}
Sch{\"a}fer,~T.; Irmler,~A.; Gallo,~A.; Gr{\"u}neis,~A. Understanding discrepancies in noncovalent interaction energies from wavefunction theories for large molecules. \emph{Nat. Commun.} \textbf{2025}, \emph{16}, 9108, DOI: \doi{10.1038/s41467-025-64104-8}\relax
\mciteBstWouldAddEndPuncttrue
\mciteSetBstMidEndSepPunct{\mcitedefaultmidpunct}
{\mcitedefaultendpunct}{\mcitedefaultseppunct}\relax
\EndOfBibitem
\bibitem[Villot \latin{et~al.}(2022)Villot, Ballesteros, Wang, and Lao]{villot2022coupled}
Villot,~C.; Ballesteros,~F.; Wang,~D.; Lao,~K.~U. Coupled cluster benchmarking of large noncovalent complexes in L7 and S12L as well as the C60 dimer, DNA--ellipticine, and HIV--indinavir. \emph{J. Phys. Chem. A} \textbf{2022}, \emph{126}, 4326--4341, DOI: \doi{10.1021/acs.jpca.2c01421}\relax
\mciteBstWouldAddEndPuncttrue
\mciteSetBstMidEndSepPunct{\mcitedefaultmidpunct}
{\mcitedefaultendpunct}{\mcitedefaultseppunct}\relax
\EndOfBibitem
\bibitem[Hansen \latin{et~al.}(2025)Hansen, Knowles, and Werner]{hansen2025accurate}
Hansen,~A.; Knowles,~P.~J.; Werner,~H.-J. Accurate calculation of noncovalent interactions using {PNO-LCCSD(T)-F12} in Molpro. \emph{J. Phys. Chem. A} \textbf{2025}, \emph{129}, 4812--4833, DOI: \doi{10.1021/acs.jpca.5c02316}\relax
\mciteBstWouldAddEndPuncttrue
\mciteSetBstMidEndSepPunct{\mcitedefaultmidpunct}
{\mcitedefaultendpunct}{\mcitedefaultseppunct}\relax
\EndOfBibitem
\bibitem[Lao(2024)]{lao2024canonical}
Lao,~K.~U. Canonical coupled cluster binding benchmark for nanoscale noncovalent complexes at the hundred-atom scale. \emph{J. Chem. Phys.} \textbf{2024}, \emph{161}, 234103, DOI: \doi{10.1063/5.0242359}\relax
\mciteBstWouldAddEndPuncttrue
\mciteSetBstMidEndSepPunct{\mcitedefaultmidpunct}
{\mcitedefaultendpunct}{\mcitedefaultseppunct}\relax
\EndOfBibitem
\bibitem[Benali \latin{et~al.}(2020)Benali, Shin, and Heinonen]{benali2020quantum}
Benali,~A.; Shin,~H.; Heinonen,~O. Quantum Monte Carlo benchmarking of large noncovalent complexes in the L7 benchmark set. \emph{J. Chem. Phys.} \textbf{2020}, \emph{153}, 194113, DOI: \doi{10.1063/5.0026275}\relax
\mciteBstWouldAddEndPuncttrue
\mciteSetBstMidEndSepPunct{\mcitedefaultmidpunct}
{\mcitedefaultendpunct}{\mcitedefaultseppunct}\relax
\EndOfBibitem
\bibitem[Shi \latin{et~al.}(2025)Shi, Della~Pia, Al-Hamdani, Michaelides, Alf{\`e}, and Zen]{Zen2025SystematicDiscrepanciesS66}
Shi,~B.~X.; Della~Pia,~F.; Al-Hamdani,~Y.~S.; Michaelides,~A.; Alf{\`e},~D.; Zen,~A. Systematic discrepancies between reference methods for noncovalent interactions within the S66 dataset. \emph{J. Chem. Phys.} \textbf{2025}, \emph{162}, 144107, DOI: \doi{10.1063/5.0254021}, Published online 2025-04-09\relax
\mciteBstWouldAddEndPuncttrue
\mciteSetBstMidEndSepPunct{\mcitedefaultmidpunct}
{\mcitedefaultendpunct}{\mcitedefaultseppunct}\relax
\EndOfBibitem
\bibitem[Awasthi \latin{et~al.}(2026)Awasthi, Otis, Huang, and Shee]{Shee2025NoncovalentAFQMC}
Awasthi,~D.; Otis,~L.; Huang,~E.; Shee,~J. Noncovalent Interaction Energies with Phaseless Auxiliary-Field Quantum Monte Carlo. \emph{J. Chem. Theory Comput.} \textbf{2026}, \emph{22}, 264--275, DOI: \doi{10.1021/acs.jctc.5c01477}, Published online 2025-12-16\relax
\mciteBstWouldAddEndPuncttrue
\mciteSetBstMidEndSepPunct{\mcitedefaultmidpunct}
{\mcitedefaultendpunct}{\mcitedefaultseppunct}\relax
\EndOfBibitem
\bibitem[Fanta \latin{et~al.}(2025)Fanta, Jure{\v c}ka, and Dubeck{\'y}]{Fanta2025NondynamicCorrelationPiPi}
Fanta,~R.; Jure{\v c}ka,~P.; Dubeck{\'y},~M. Why Nondynamic Correlation Matters for {$\pi\pi$} Stacking? {L}essons from the Benzene Dimer. \emph{J. Phys. Chem. Lett.} \textbf{2025}, \emph{16}, 10982--10988, DOI: \doi{10.1021/acs.jpclett.5c02576}\relax
\mciteBstWouldAddEndPuncttrue
\mciteSetBstMidEndSepPunct{\mcitedefaultmidpunct}
{\mcitedefaultendpunct}{\mcitedefaultseppunct}\relax
\EndOfBibitem
\bibitem[Lambie \latin{et~al.}(2026)Lambie, L{\'o}pez-R{\'i}os, Kats, and Alavi]{LambieAlavi2025AcOHdimer}
Lambie,~S.; L{\'o}pez-R{\'i}os,~P.; Kats,~D.; Alavi,~A. Nodal error behind discrepancies between coupled cluster and diffusion Monte Carlo in hydrogen-bonded systems. 2026; \url{https://arxiv.org/abs/2508.17937}, arXiv:2508.17937v2 [physics.chem-ph], submitted 2026-01-21\relax
\mciteBstWouldAddEndPuncttrue
\mciteSetBstMidEndSepPunct{\mcitedefaultmidpunct}
{\mcitedefaultendpunct}{\mcitedefaultseppunct}\relax
\EndOfBibitem
\end{mcitethebibliography}

\end{document}